# Towards a probabilistic foundation of relativistic quantum theory: the one-body Born rule in curved spacetime

**Maik Reddiger** · **Bill Poirier**

Received: 3 April 2024 / Accepted: 12 August 2024 / Published online: 30 October 2024

**Abstract** In this work, we establish a novel approach to the foundations of relativistic quantum theory, which is based on generalizing the quantum-mechanical Born rule for determining particle position probabilities to curved spacetime. A principal motivator for this research has been to overcome internal mathematical problems of relativistic quantum field theory (QFT) such as the 'problem of infinities' (renormalization), which axiomatic approaches to QFT have shown to be not only of mathematical but also of conceptual nature. The approach presented here is probabilistic by construction, can accommodate a wide array of dynamical models, does not rely on the symmetries of Minkowski spacetime, and respects the general principle of relativity. In the analytical part of this work, we consider the 1-body case under the assumption of smoothness of the mathematical quantities involved. This is identified as a special case of the theory of the general-relativistic continuity equation. While related approaches to the relativistic generalization of the Born rule assume the hypersurfaces of interest to be spacelike and the spacetime to be globally hyperbolic, we employ prior contributions by C. Eckart and J. Ehlers to show that the former condition is naturally replaced by a transversality condition and that the latter one is obsolete. We discuss two distinct formulations of the 1-body case, which, borrowing terminology from the non-relativistic analog, we term the Lagrangian and Eulerian pictures. We provide a comprehensive treatment of both. The main contribution of this work to the mathematical physics literature is the development of the Lagrangian picture. The Langrangian picture shows how one can address the 'problem of time' in this approach and, therefore, serves as a blueprint for the generalization to many bodies and the case that the number of bodies is not conserved. We also provide an example to illustrate how this approach can in principle be employed to model particle creation and annihilation.

M. Reddiger (✉)
Department of Computer Science and Languages, Anhalt University of Applied Sciences, Lohmannstraße 23, 06366 Köthen, Germany
e-mail: maik.reddiger@hs-anhalt.de

B. Poirier
Department of Chemistry, University of Vermont, E200 Innovation Building, Burlington, VT 05405, USA
e-mail: bill.poirier@uvm.edu

M. Reddiger · B. Poirier
Department of Physics and Astronomy, Texas Tech University, Box 41061, Lubbock, TX 79409-1061, USA

M. Reddiger · B. Poirier
Department of Chemistry and Biochemistry, Texas Tech University, Box 41061, Lubbock, TX 79409-1061, USA

## Contents

## 1 Introduction

To support the reader in navigating this work, we begin with a brief outline of this section.

The general motivation for this research in the context of the history of modern particle physics is provided in Sect. 1.1. In Sect. 1.2, we argue why the (general-)relativistic generalization of the Born rule provides a natural approach to the foundations of relativistic quantum theory. Respective analytical results on the one-body theory, which constitute the core contribution of this work, are summarized in Sect. 1.3. In Sect. 1.4 we provide some remarks on the directly related literature. Notations and conventions are given in Sect. 1.5.

Readers only interested in our analytical results are invited to skip to Sects. 1.3 and 1.5.

### 1.1 Historical background and motivation

The Standard Model of Particle Physics (SM) is contemporarily the empirically most successful theory of the fundamental constituents of matter and its mutual interactions. While it does not incorporate the gravitational interaction, in its ability to provide such a unified description lies arguably its greatest achievement.[1]

[1] References [10,220] provide elementary introductions to the SM for readers lacking a background in particle physics. A more in depth treatment is given in Ref. [186]. Part II of Ref. [221] as well as Ref. [176] constitute differential-geometric accounts.

The theoretical framework that underlies the SM is known as relativistic quantum field theory (QFT). Historically, the abstract framework of QFT emerged out of perturbative descriptions of physical phenomena involving electrons and electromagnetic radiation: quantum electrodynamics (QED) in its early stages. First articles on the topic appeared in the late 1920s and early 1930s [121].[2] In order to attain a physically acceptable theory of radiation that is also able to describe quantum phenomena of matter, one ultimately had to generalize the non-relativistic theory of quantum mechanics (QM) to the (special-)relativistic setting.[3] In this generalization lies the theoretical justification for the existence of QFT (cf. p. xi in Ref. [214]).

From the 1930s up until the late 1940s QED was, however, generally viewed with great skepticism. This was mostly due to the appearance of unphysical divergences in the theory, the so called 'problem of infinities' (cf. Sec. 1.3 in Ref. [214], Ref. [121], Sec. 5 in Ref. [36], and Ref. [178]). It was the perceived failure of the 1-electron Dirac theory [46] to account for a variety of spectroscopic measurements [70,123,124,149,150], that created the need for an alternative approach. Bethe's and other theoreticians' success in accounting for those experimental anomalies within QED enabled the theory to meet this demand.[4] While their method of 'renormalizing' the charge and mass of the electron was ad hoc, mathematically unsound, and it did not entirely resolve 'the problem of infinites' [52,71], the lack of empirically viable competition meant that QED was to emerge victoriously.

Moreover, the success of QED in this question was going to determine subsequent developments in theoretical particle physics. QFT became the language of the field moving forward, thus ingraining the idea of *renormalization*. Despite overt mathematical problems of the formalism, over the coming decades QFT proved itself to be sufficiently flexible to accommodate and sometimes even predict new experimental data.[5] Undoubtedly, it was also the close collaboration between theoreticians and experimentalists during that time that eventually enabled QFT to serve as the theoretical foundation for the SM we employ today.[6]

While the SM represents a significant milestone in the history of particle physics, there is a growing concern with regards to the problem that the vast majority of so called 'Beyond the Standard Model' research has provided few, if any, empirically tangible results [104,105,193,219].[7] In this respect, it is worth noting that in the history of science, it is not uncommon for a subject area to pass through such a phase of apparent stagnation. Overcoming it in a scientifically fruitful manner usually necessitates a careful reevaluation of underlying methodologies and assumptions. Although this process is commonly resisted by practitioners in the subject, it is nonetheless necessary for its scientific integrity (cf. Refs. [122,158]).

In order to address at least some of the internal mathematical problems of QFT,[8] mathematical physicists began the process of crafting it into a rigorous formalism in the early 1950s. Starting with works by Friedrichs [76–80], Wightman [216,217], Cook [37] and others [5,114,130,183], this methodological shift eventually gave rise to a variety of research programs and axiomatic systems: the most well-known ones are 'axiomatic quantum field theory' (the Wightman axioms; cf. Ref. [217] and Sec. 3.1 in Ref. [194]), 'algebraic quantum field theory' (the Haag–Kastler axioms; cf. Refs. [87,88]), and 'constructive quantum field theory' (the Osterwalder-Schrader axioms and extensions by Glimm and Jaffe; cf. Refs. [85,156,157]). Though axiomatic approaches to QFT were at times being criticized for their limited practical applicability (cf. Sec. II in Ref. [32]), today their scientific value has become generally acknowledged.[9] Axiomatic approaches to QFT have contributed to particle physics by clarifying both the concepts and mathematical structures employed in 'perturbative QFT' [71].

---

[2] See Refs. [25,44,63,64,95,96,111–113,209,215], as well as the early reviews [65,155].

[3] Recall that Maxwell's electromagnetism is strictly speaking a special-relativistic theory. The physical need to account for relativistic effects arises out of the empirical untenability of the Newtonian conception of space and time, even in the absence of gravity (e.g. time dilation and length contraction). For point masses this difference becomes noticeable at high velocities relative to the chosen frame of reference.

[4] See [18,29,30,53,67,75,120,187,188,202,203] for original references.

[5] The detection of the Higgs boson in 2012 is perhaps the most famous example.

[6] See Refs. [118,119]. We also refer to Ref. [31] for a more detailed history of the SM.

[7] See also p. xvi in Ref. [186], Chap. 11 in Ref. [33] and p. 30 sq. in Ref. [31].

[8] The article by Fraser [71] provides a general overview of the most important problems.

While axiomatic approaches to QFT constitute an important methodological step in the history of theoretical particle physics, we wish to articulate the following objection to purely formal approaches to resolving the difficulties of perturbative QFT: The search for an axiomatic foundation for the subject is necessarily accompanied by an identification of its basic concepts and how those relate to the natural phenomena one intends to explain. Especially with regards to the concept of renormalization,[10] it is worthwhile to remember that the goal is not to reproduce what is known to be inconsistent. Rather, it is to construct an internally consistent relativistic quantum theory that can compete with the empirical successes of the SM.

Achieving this goal may require a revision of the basic concepts of contemporary relativistic quantum theory. Although this may seem like a bold task at first, one may find hope in recalling that QFT is merely the language in which developments of the field have historically been formulated [118]—a language that is still questionable on logical grounds.[11] To acknowledge the need to change this language does not necessarily translate to 'throwing out the baby with the bathwater'. It merely means acknowledging the need for a different theoretical perspective on existing results.

Providing such a perspective is indeed the primary aim of this work and the research program it is intended to give rise to.[12]

### 1.2 The Born rule as a novel approach to relativistic quantum theory

The approach suggested and pursued in this work is based on a generalization of the Born rule for position in (non-relativistic) QM.

Following Pauli's formulation [14,159], the rule determines the probability that a given number of bodies $N \in \mathbb{N}$ is located in a particular 'region' of configuration space at a given time. More precisely, if $\rho(t, \vec{x}_1, \ldots, \vec{x}_N)$ denotes the values of the (time-dependent) probability density $\rho$ at time $t \in \mathbb{R}$ and at points $(\vec{x}_1, \ldots, \vec{x}_N) \in \mathbb{R}^{3N}$,[13] as

[9] This is perhaps best exemplified by their explicit mention in the official problem description for one of the so called 'Millenium Prize Problems' [108]. See also Ref. [71].

[10] Notable historical critics of the idea of renormalization include, but are not limited to, Paul Dirac [43], Lev Landau, and Gunnar Källén (cf. Ref. [213] and p. xi sqq. in Ref. [115]). Dirac had the following to say on the matter [43]:

> It turns out that, sometimes, one gets very good agreement with experiments working with these rules. In particular, if one has charged particles interacting with the electromagnetic field, these rules of renormalization give surprisingly, excessively good agreement with experiments. Most physicists say that these working rules are, therefore, correct. I feel that is not an adequate reason. Just because the results happen to be in agreement with observation does not prove that one's theory is correct. After all, the Bohr theory was correct in simple cases. It gave very good answers, but still the Bohr theory had the *wrong concepts*. [emphasis added]

[11] See Sec. 2.6 in Ref. [170] for a more in depth elaboration on this point with regards to the problem of divergences. In the literature, Haag's theorem is also commonly put forward as a foundational objection towards QFT, see e.g. Refs. [6,54,71,72,116,141,190]. A criticism that does not seem to have received much attention in the literature so far is the ad hoc nature of quantization (cf. Footnote 21 below).

[12] We believe that in this respect Segal's words have kept their relevance (cf. p. 469 in Ref. [189]):

> It seems that for foundational purposes only a quite comprehensive attack employing conservative but global methods has much hope of ultimate success. As this has never really precisely been undertaken, there is no reason for undue pessimism, but the scope of such a development is necessarily such that it is unrealistic to begin highly explicit analytical computations until the fundamental design is well established.

[13] Of course, for each $t$ the function $\rho(t, \,.\,)$ is only determined up to a set of (Lebesgue) measure zero in $\mathbb{R}^{3N}$. But this is only of minor relevance to the discussion here.

obtained from a respective (normalized) $N$-body Schrödinger [185] or Pauli [160] wave function, then

$$\mathbb{P}_t(U) = \int_U \rho(t, \vec{x}_1, \ldots, \vec{x}_N)\, \mathrm{d}^3x_1 \ldots \mathrm{d}^3x_N \tag{1.1}$$

gives the probability that at time $t$ the bodies are located at a configuration $(\vec{x}_1, \ldots, \vec{x}_N)$ within the (Lebesgue) subset $U$ in $\mathbb{R}^{3N}$.

*Remark 1.1* Since the Born rule is tied to the projection postulate via the axioms of QM,[14] we shall comment upon the latter in the context of this work. Thereby, we also address some related questions of interpretation.

Above all, here we do not consider the question of if and how the projection postulate ought to be implemented in relativistic quantum theory.[15] Indeed, due to the connection to the problem of "measurement", the status of the projection postulate has been a subject of scientific debate also in the context of the non-relativistic theory (see e.g. Refs. [2,7,13,60,62,161,206]). Since it is beyond the scope of this work to address this point, we have chosen a formulation that aims to be as neutral as possible with regards to the usage of the term "measurement". Readers preferring the traditional interpretation may understand the terminology that the bodies "are located" in a certain region in the "active" sense that they are "located by measurement", whereas the terminology also allows for an interpretation in the "passive" sense that their objective positions lie in the region of interest (as in, for instance, de Broglie-Bohm theory [21,99] and related approaches [62,152,164]).

Indeed, the theory we develop in this work is most easily understood from the point of view of the statistical/ensemble interpretation of QM. In essence, this interpretation states that the primary utility of QM is to make probabilistic predictions on the "physical observables" of "similarly prepared systems" of particles—that is the *ensemble* obtained by taking the collection of all *samples*.[16]

While one would be justified to view this as a "minimalist" interpretation of QM, there does exist an implicit conflict with the Copenhagen interpretation insofar as statements such as "the electron is in state Ψ" become meaningless—the words "state" and "quantum system" always refer to an ensemble of physical (sample) systems, not the systems themselves. That the ensemble view is compatible both with the aforementioned "active" and "passive" interpretations was clarified by Home and Whitaker [102], leading to the distinction between "minimal" and "pre-assigned initial values" ensemble interpretations, respectively.[17] For a general introduction to the subject the reader is referred to Refs. [8], [161], and Sec. 2.1 in Ref. [170]. ◇

For the purpose of constructing a relativistic quantum theory, there are two major reasons why one would focus on this particular aspect of the non-relativistic theory:

First, as opposed to vectors in an abstract Hilbert space or even elements of the Fock space constructed thereof, the above Born rule is formulated directly in terms of spatio-temporal concepts. One may therefore expect there to be a more or less straightforward generalization to the relativistic setting.

---

14 Dropping the explicit time-dependence for simplicity, the precise connection is as follows:
For a quantum observable $\hat{A}$ on the Hilbert space $(\mathcal{H}, \langle\,.\,,\,.\,\rangle)$ consider the corresponding projection-valued measure $I \mapsto \mathrm{P}_I$ on the Borel sets $\mathcal{B}(\mathbb{R})$, as defined via the functional calculus (cf. Sec. VIII.3 in Ref. [173]). Further let $\Psi \in \mathcal{H}$ be normalized and for $I \in \mathcal{B}(\mathbb{R})$ with $\mathrm{P}_I\,\Psi \neq 0$ set $\mathrm{R}_I\,\Psi = \mathrm{P}_I\,\Psi / \|\mathrm{P}_I\,\Psi\|$. Then the Born rule [26–28] states, that the quantity $|\langle \mathrm{R}_I\,\Psi, \Psi\rangle|^2$ gives the probability that upon "measuring" $\hat{A}$ one obtains values in $I$. If values in $I$ are indeed "measured", then the projection postulate demands that $\mathrm{R}_I\,\Psi$ is the state of the system "after measurement".
At least heuristically, Eq. (1.1) (for fixed $t$) is derived from the fact that the scalar position observables commute in some appropriate sense and are therefore "simultaneously measurable", so that the corresponding projections can be unambiguously combined. More specifically, one denotes the characteristic function for a corresponding (Borel or Lebesgue) set $U$ by $\chi_U$ and sets $\mathrm{P}_U\,\Psi = \langle \chi_U \Psi, \Psi\rangle\, \chi_U \Psi$ to obtain the probability in Eq. (1.1) via the above Born rule.

15 There do exist works that attempt to implement the projection postulate in the relativistic context, see e.g. Refs. [66,132,195].

16 An ensemble is a theoretical abstraction, which is formally taken to be the (uncountable) set of all samples of interest. As in statistical mechanics, one can define an ensemble by considering, for instance, all samples with a given energy.

17 The former belong to the more "orthodox" interpretations of QM, while the latter may be viewed as "hidden variable" interpretations.

Second, the Born rule is fundamentally a probabilistic statement: Taking its generalization as a point of departure for the relativistic theory guarantees that the latter is of probabilistic nature from the onset.

Evidently, this will also tie the particle concept into the basic structure of the theory. Yet, this is should not come as a surprise, since the particle concept is central to the non-relativistic theory[18] and any straightforward relativistic generalization will therefore inherit it.[19,20]

In their 1929 article [95], which was foundational to the development of quantum field theory, Heisenberg and Pauli did indeed consider a particle-based approach. They chose to reject it on the following grounds:

> As is generally known, in classical point mechanics a relativistically invariant formulation of the many-body problem with the aid of the Hamiltonian theory is not feasible. One may therefore not hope that in the quantum theory a relativistically invariant treatment of the many-body problems with differential equations in configuration space […] will be attainable […]
> [translated from German]

Heisenberg and Pauli did thus not consider the notion that a particle-based relativistic quantum theory could be constructed which does not rely on 'quantizing' a 'classical' Hamiltonian system.[21]

This is the point at which this work parts ways with their reasoning.

Rest assured, we do not deny that there may be physical situations in which the concept of a discrete particle ceases to be a viable physical concept and a field description becomes more appropriate. Yet due to the centrality of the particle concept to QM and its non-relativistic siblings (see e.g. Refs. [49,152,169,172]), such situations are beyond the purview of a mere relativistic generalization of those theories. The conservative approach is therefore to put the particle concept at the center of such a generalization—as Heisenberg and Pauli implicitly admitted with their above statement.

Still, the historical context justifies the assertion that the construction of a relativistic quantum theory with the particle concept at its center has shown itself to be a difficult problem.

In particular, if one does take the generalization of the Born rule as an approach to the problem, there are two major hurdles one needs to address:

(1) The so-called 'problem of time'[22]: On one hand, Eq. (1.1) above indirectly refers to the different spatial positions of the bodies at the same instance of time. On the other hand, in the theory of relativity the notion of simultaneity lacks the physical significance it has in non-relativistic theories.[23]
In special relativity this is known to be a consequence of the special principle of relativity in conjunction with the invariance of the speed of light (in vacuum). In the general theory of relativity, it may be considered a consequence of the *general principle of relativity*. Einstein [61] phrased the latter as follows:

---

[18] In order to avoid any deeper discussions on the interpretation of quantum-mechanical wave functions here, we merely point out that the number of bodies $N$ is a central ingredient in devising a quantum-mechanical description for a physical system (even in the case that the asymptotic limit $N \to \infty$ is the one of interest).

[19] Though an in-depth discussion of how particle creation and annihilation is handled in the formalism is beyond the scope of this work, Sect. 4 provides a simple example to illustrate the general idea of how to do so.

[20] Common objections to particle-based relativistic quantum theories in the modern literature [93,138] are discussed at the end of Sect. 1.4 below.

[21] In Sec. 1 of Ref. [169], the first author of this work provided a detailed critique to employing the concept of 'first quantization' as a tool of scientific theorizing in quantum theory.
The criticism does, however, not entirely carry over to the concept of 'second quantization' or 'field quantization', as it is commonly referred to. While the idea of obtaining a quantum theory from an ad hoc modification of a 'classical theory' is also problematic in this instance for essentially the same reasons, the introduction of Fock spaces [69] was a necessary step to move from 1-body theories to theories that are able to account for a varying particle number (cf. p. 190 in Ref. [96]). The interested reader is referred to Ref. [37].

[22] Refs. [4,106] give a general introduction to the problem. While the authors view it as a problem of 'quantum gravity', it is arguably a problem to be addressed in any relativistic quantum theory—even those that take Minkowski spacetime as a 'fixed background'.

[23] The philosopher of science Hans Reichenbach famously argued on the basis of relativity theory that simultaneity is a matter of convention [174,175].

We shall be true to the principle of relativity in its broadest sense if we give such a form to the laws that they are valid in every such four-dimensional system of co-ordinates, that is, if the equations expressing the laws are co-variant with respect to arbitrary transformations.

If understood in a broad sense, the general principle of relativity also demands that we may not employ any quantities in the formulation of fundamental physical laws that depend on a particular choice of physical observer, frame of reference, 'initial' hypersurface, etc.

A priori, there is thus reasonable doubt as to whether a relativistic generalization of the Born rule can be made sense of. It is indeed a matter of simple counting that in an $(n+1)$-dimensional spacetime $\mathcal{Q}$ with $N$ bodies ($n, N \in \mathbb{N}$) the most obvious candidate for a 'relativistic configuration space', the $N$-fold product manifold $\mathcal{Q}^N$, is $N(n+1)$-dimensional—so that there are $N-1$ 'too many time dimensions'.

(2) The problem of dynamics: For Eq. (1.1) to define a probability measure at each $t \in \mathbb{R}$, the probability conservation law $\mathbb{P}_t(\mathbb{R}^{3N}) \equiv 1$ needs to hold. In non-relativistic QM this is assured by the unitarity of the time evolution operator. In particular, for time-independent Hamiltonians $\hat{H}$ unitarity of $U_t = \exp(-\imath t \hat{H}/\hbar)$ for each $t$ follows from self-adjointness of $\hat{H}$. This, in turn, suggests that for the relativistic theory we need to introduce a (possibly indirect) assertion on how the time evolution is modeled. Yet this makes it difficult to fully separate the probabilistic and the dynamical aspects of the theory.

Historically, the conceptual problems with viewing the Dirac equation as a 1-body evolution equation were indeed one of the main motivators for pursuing the development of quantum field theory over a deeper understanding of relativistic $N$-body theories.[24]

There are two main works in the recent literature that propose a relativistic generalization of the Born rule:

In Ref. [132], Lienert and Tumulka suggest a construction for $N$ bodies in Minkowski spacetime which takes the configuration space to be the $N$-fold product of a Cauchy surface therein.[25] In this work, an 'instant of time' is viewed as a choice of Cauchy surface. Different Hilbert spaces are associated with the respective hypersurfaces and the existence of a unitary time-evolution operator between the two surfaces is postulated, in analogy to the Schrödinger picture of QM. The authors do not commit to any particular dynamical models, though the considered examples were all descendants of the (1-body) Dirac equation (cf. Sec. 4 in Ref. [132]). For the many-body case, the considered models seem to lack physical justification, with the exception of "free Dirac evolution".[26]

Contrarily, Miller et al. [144] consider the more general case of a globally hyperbolic spacetime and then use Bernal and Sanchez' stronger version [15,16] of Geroch's Splitting Theorem [84] to construct an '$N$-particle configuration spacetime'. The latter is chosen to be the product of $\mathbb{R}$ with the $N$-fold product of a Cauchy surface (cf. Sec. 2.2 in Ref. [144]). The authors were indeed able to prove an invariance theorem to address the 'problem of time' (cf. Thm. 22.(ii) in Ref. [144] as well as Thm. 2 in Ref. [143]). Yet the proposed dynamical equations violate the general principle of relativity (cf. Equations 20 and 24 in Ref. [144]).

In this work, we develop the theory of the 1-body Born rule on curved spacetime under the assumption of smoothness of the mathematical quantities involved. This novel generalization mainly draws from prior contributions to the theory of general-relativistic fluid mechanics due to Eckart [55] and Ehlers [59]. We show that there are two distinct formulations of the theory for the case that one allows for the temporal evolution of the quantities involved. In full analogy to the non-relativistic analog, we term those formulations the *Lagrangian picture* and the *Eulerian picture*, respectively.[27]

---

[24] Heisenberg and Pauli began their 1929 article [95] by stating that the problem of how to treat radiation in quantum theory had not been entirely resolved and that a relativistically invariant formalism was needed to adequately describe light-matter interaction. They went on to write that "[t]his problem seems to be fundamentally connected with the great difficulties that according to Dirac obstruct a relativistically invariant formulation of the one-electron problem, and one will only attain a fully satisfactory solution of the task assigned here after a clarification of those fundamental difficulties." [translated from German]

[25] See also Ref. [133].

[26] In particular, those models do not explicitly account for the general principle of relativity. In Rem. 6 of Ref. [132], Lienert and Tumulka indeed discuss so-called 'Hypersurface Bohm-Dirac models', which intentionally violate the principle (see also §11 in Ref. [48] and Sec. I in Ref. [51]). Following Dürr et al. (cf. Chap. 9 in Ref. [49]), this approach had been developed in various prior works [17,48,51,180] in an attempt to find a relativistic theory of Bohmian mechanics. It goes back to one of the authors' own articles (cf. §8 to §12 in Ref. [48]) and works composed by Bohm and Hiley (cf. Ref. [24], as well as §10.4 and §10.5 in [21]). Though it is possible to formulate such theories within the mathematical formalism of general relativity, the violation of one of its core principles deserves physical justification.

The development of the Lagrangian picture we view as the main contribution of this work to the foundations of relativistic quantum theory: The construction of the Lagrangian picture for the 1-body case opens up a potentially viable path towards the generalization of the theory to the $N$-body case or even the case that the number of bodies is not conserved.

Like Miller et al. [143,144], we address the 'problem of time' in this formulation by proving an invariance property that assures that the general principle of relativity is indeed respected (cf. Theorem 3.1.(4) and Corollary 3.1.(4)).[28]

While we do not suggest any explicit dynamical models, we show in several examples how the Dirac equation fits into the formalism. The structural aspects of the theory are the focus of this work. We thereby provide a mathematical framework that is in principle neutral with respect to the dynamics one wishes to impose—be it through a relativistic wave equation or other (possibly non-linear) partial differential equations for the respective fundamental quantities of the theory.

The general idea that the probabilistic and the dynamical aspects of relativistic quantum theory may be treated separately is inspired by the non-relativistic Madelung equations [135,136] and their close connection to Kolmogorov's theory of probability (cf. Chap. 2 in Ref. [170] and Ref. [168]). Though their precise mathematical relation to the Schrödinger equation is still a subject of mathematical research [34,82,140,169,172,210–212], the Madelung equations allow one to separate the equation of continuity from 'the dynamics', as encoded by an equation containing the forces/potentials.

In the relativistic theory, one is thus not bound to the difficult task of guessing appropriate wave equations that have to make sense in a probabilistic context—as Dirac was miraculously able to do [41,46].

The reason that we have pursued a generalization of the Born rule to the general-relativistic setting, as opposed to being satisfied with a special-relativistic version, is that this approach is not only more general but simpler: As Hollands and Wald have noted on p. 87 sq. of their article [100], Minkowski spacetime constitutes a highly symmetrical setting, yet even a special-relativistic quantum theory ought not to rely on those symmetries in its basic formulation. From a mathematical perspective those symmetries constitute 'obsolete assumptions', thus making a construction of the theory on curved spacetime the natural approach.[29] Of course, for practical purposes and at this point in time, one is nonetheless primarily interested in how the theory works in Minkowski spacetime.

Ultimately and as explained in Sect. 1.1 above, the mathematical axiomatization of relativistic quantum theory requires an identification of its basic concepts. It is in this sense that the progress that was achieved in the foundations of QM via the development of the ensemble interpretation has a direct bearing on the theory we develop here. Yet since the work is primarily of mathematical nature, we use the ensemble interpretation mostly for methodological purposes (cf. [191]). The reader is, of course, free to choose their own interpretation. Still, how one resolves the fundamental questions of the non-relativistic theory is bound to lead to different judgments on whether one deems a particular approach to the relativistic theory as reasonable or not. That is, the ongoing debate on the conceptual foundations of quantum mechanics [73,74,109,110] cannot be fully separated from the discussion of the mathematical foundations of relativistic quantum theory.

[27] We refer to Refs. [98,164,199] for a discussion of how the Langrangian picture fits into non-relativistic quantum theory.

[28] Strictly speaking, it is the ability to freely choose the 'initial' hypersurface (under general suitable conditions) that assures that the general principle of relativity is indeed respected. The invariance property only holds in case one has probability conservation—which is physically sensible.

[29] We refer to Refs. [9,100,101] for an introduction to 'Quantum Field Theory on Curved Spacetime'. It should be noted, however, that we pursue a different approach here. In general, a 'quantum theory on curved spacetime' is needed whenever one is in the relativistic regime and the gravitational field is strong enough to have a noticeable impact on the dynamics of the quantum system. It ought to be valid as long as the influence of the gravitational field of the quantum system on its own dynamics is negligible. Examples of such systems include an atom in the vicinity of a small black hole and a chemical reaction of two molecules under the influence of a strong gravitational wave.

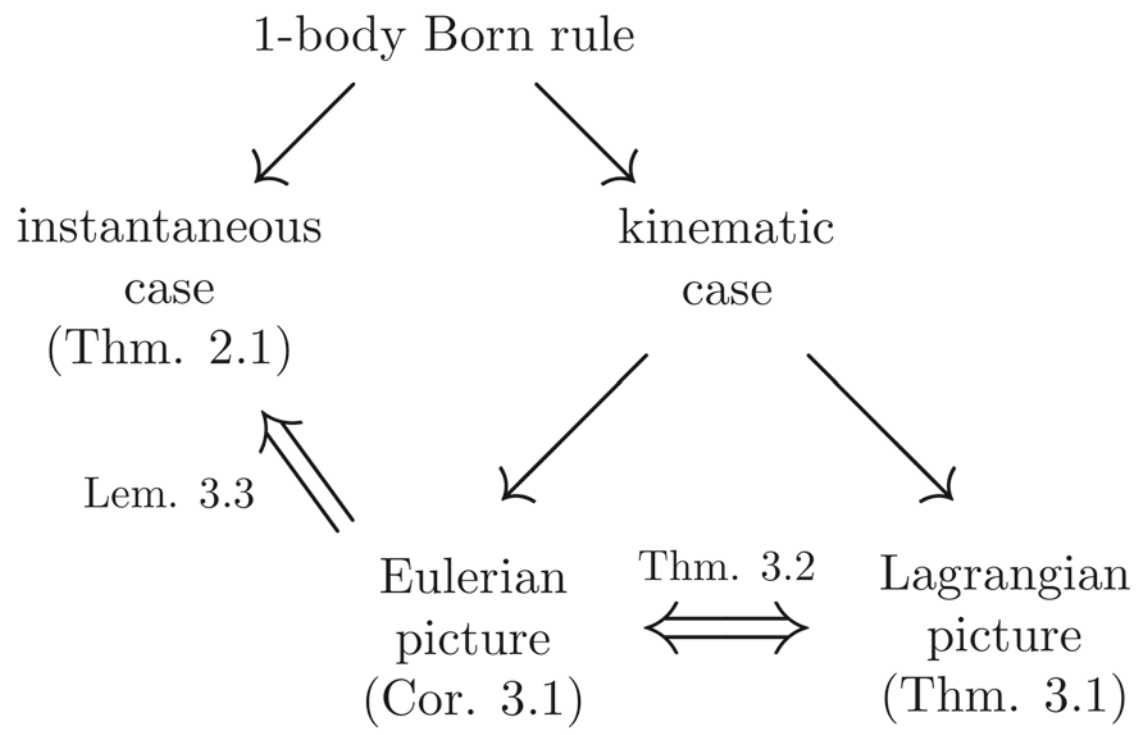

**Fig. 1** The sketch shows the general structure of the theory laid out in this work along with the respective main theorems. A more detailed outline is given in Sect. 1.3

## 1.3 Summary of analytical results and outlook

The central contribution of this work is the comprehensive development of the theory of the 1-body Born rule on curved spacetime under the assumption of smoothness of the mathematical quantities involved.[30]

From a mathematical perspective, this theory may be viewed as a special case of the theory of the general-relativistic continuity equation. Indeed, if one considers the non-relativistic analog, this is to be expected: As long as the corresponding scalar quantity of interest is obtained from an integral of a scalar density over space, it matters little to the mathematical formalism whether that integral determines a mass, a charge, or a probability. Therefore, one reobtains the generic case by replacing the normalization condition on the density by the condition of Lebesgue integrability, (see also Remark 1.2 below). Accordingly, much of the prior progress this work builds on was achieved in other areas of the general theory of relativity [55,59]. The reader is referred to Sect. 1.4 below.

With the aforementioned restrictions, this work is the first to consider the theory of the general-relativistic continuity equation in full generality. As in non-relativistic continuum dynamics, we find that there are two different formulations of the theory: The *Lagrangian picture* and the *Eulerian picture*. While various discussions of the Eulerian picture may be found in the literature (see e.g. §3.0.2 in Ref. [179], p. 50 sq. in Ref. [59], and p. 69 sq. in Ref. [92]), an in-depth discussion of the Lagrangian picture—and thus the need to distinguish between the two[31]—constitutes a novel contribution of this work. For the Eulerian picture we introduce the transversality condition and use it to show novel theorems, which overcome the overly restrictive conditions of spacelikeness of the 'initial' hypersurface and global hyperbolicity of the spacetime that are common in the related literature (cf. Refs. [132,144] and Sect. 1.4 below).

A general overview of the theory is given in Fig. 1. The theorems referred to therein constitute the central results of this work. We formulate them as 'self-contained packages', which comes at the expense of obtaining longer statements. The intention is to make it easy for the reader to directly jump to the important statements, referring to the (italic) definitions in the text for terminology.

The 'instantaneous case' is discussed in Sect. 2 and constitutes a treatment of the Born rule in the absence of temporal evolution (i.e. for 'fixed time'). The central result of this section is Theorem 2.1, with Proposition 2.1 providing a general existence result. Example 2.1 shows how the 1-body Dirac theory fits into the general framework developed here.

The 'kinematic case' is needed in order to account for temporal evolution. It is the topic of Sect. 3.

---

30 Do note, however, that in vicious spacetimes a more general definition of flowouts is appropriate, so that some results in Sect. 3 would have to be weakened (cf. Rem. 3.1.3) and Appendix C).

31 It should be noted that Sklarz and Horwitz used the term 'Eulerian velocity field' in their work [192], thus implicitly suggesting the existence of a Lagrangian picture in the relativistic setting.

Section 3.1 begins the treatment of the kinematic case with a general conceptual discussion. Here, we physically motivate the splitting the current density vector field $J$ into an 'invariant probability density' $\rho$ and a velocity vector field $V$ via Eq. (3.3). We show that, while the time evolution induced by $V$ does generally not preserve the condition of spacelikeness of an 'initial' hypersurface (Example 3.1), it does always preserve the transversality condition (Proposition 3.1). This is a central justification for choosing the latter condition over the former one.

The full mathematical theory for the kinematic case is developed in Sects. 3.2 and 3.3. In the former one we discuss the Lagrangian picture, for which the concept of a 1-body flowout is central (Definition 3.3). Proposition 3.2 proves existence and Lemma 3.1 establishes its utility within the general theory of relativity. The main result of this subsection is Theorem 3.1. Lemma 3.2 provides relevant coordinate expressions, and we consider a simple example in Example 3.2.

An analogous treatment of the Eulerian picture is given in Sect. 3.3. Therein we also discuss the mutual relation between the two pictures. Indeed, they may be regarded as mathematically equivalent (Theorem 3.2; justifying the corresponding arrow in Fig. 1). The central theorem on the Eulerian picture is Corollary 3.1. Though coordinate expressions are known, we do treat an example here as well (Example 3.3). To make the internal consistency of the theory explicit, we have also shown how 'at fixed time' the kinematic Born rule indeed yields the instantaneous one in Lemma 3.3 (the implication arrow in Fig. 1).

Finally, in Sect. 4, we discuss an example for which probability conservation is intentionally violated. Its purpose is not to propose any specific physical model, but to show how the theory can in principle be employed to describe the case of a varying number of bodies.

Before we address potential future developments of the theory, we shall make some remarks of a general nature.

*Remark 1.2* (1) In this work, we allow for future-directed causal current density vector fields $J$, which includes the cases that it is future-directed timelike or lightlike. While other authors have also found this 'causal evolution' to be worthy of investigation [56,143,144], we consider it an open question whether there are any fundamental physical models for which $J$ is not timelike.[32]

To introduce a conserved current of the electromagnetic field, for instance, one requires a spacetime symmetry (cf. p. 61 sqq. in Ref. [92]) or the JWKB approximation (cf. Sec. 3.1 and 3.2 in Ref. [184]), either one of which puts a strong restriction on the physical context.[33]

(2) For readers with a focus on mathematics we wish to point out the possibility of defining natural 'time-invariant' $L^p$-spaces, as indicated in Remarks 3.3 and 3.5.

(3) It is straightforward to adapt the theory here for the purposes of relativistic fluid mechanics by dropping the interpretation of $\rho$ as an invariant probability density and interpreting it as an invariant (inertial) mass density instead.[34] Of course, the respective physical dimensions of $\rho$ and $J$ also need to be changed accordingly. Interested readers are referred to the introductory discussion in Sect. 3.

It is also worthy of note that there are various works in the literature that make use of the historical, yet outdated concepts of 'relativistic mass' and 'rest mass' [1,154]. That those notions lack physical justification can be seen by requiring mass conservation (via the relativistic continuity equation) and viewing the point mass model as the limit of an underlying continuum-theoretical one—in full analogy to the non-relativistic theory. Accordingly, the concept of a 'rest mass density' is also problematic.

(4) If the reader wishes to adapt the theory here to the case that $J$ is a charge current density vector field, then it is important to keep in mind that inverting the signs of the charge inverts the direction of $J$ at each point. Because of this, it is simpler to work with the invariant charge density $\rho$ and the future-directed timelike velocity field $V$, which are related to $J$ via Eq. (3.3) below. We again refer to the introductory discussion in Sect. 3.

It is also worthwhile to rewrite the special-relativistic inhomogeneous Maxwell equations in terms of the invariant charge density, which differs by the common ‘charge density’ by a ‘$\gamma$-factor’ (cf. Equations (3.3) and (3.4) below):

$$\nabla \cdot \vec{E} = \frac{1}{\varepsilon_0} \frac{\rho}{\sqrt{1-\left(\frac{\vec{v}}{c}\right)^2}} \quad , \quad \nabla \times \vec{B} = \mu_0 \frac{\rho\,\vec{v}}{\sqrt{1-\left(\frac{\vec{v}}{c}\right)^2}} + \frac{1}{c^2}\frac{\partial \vec{E}}{\partial t}. \tag{1.2}$$

[32] The requirement that the current density should be nowhere spacelike constitutes a causality condition imposed on the physical theory. Therefore, the Klein-Gordon equation cannot serve as a physical evolution equation [83].

[33] We also refer to Synge's critical account [197] on the topic of 'photon wave functions'.

[34] We use the word 'inertial' here to separate it from to concept of 'gravitational mass'. In general relativity, the latter is always tied to the metric while the former need not be (under suitable approximations).

Here, we used SI units and notation that is standard in the physics literature.

◇

The relativistic generalization of the Born rule for the 1-body case constitutes a prerequisite for the generalization to $N$ bodies. The latter poses a conceptually non-trivial step, since statistical correlation and the need to account for the indistinguishability of bodies only become relevant for $N > 1$. Nonetheless, it is indeed possible to generalize the Lagrangian picture of the 1-body theory to the $N$-body case. Roughly speaking and without going into any detail here, this makes it possible to introduce a single ‘time parameter’ without violating the general principle of relativity. Developing this theory in its full scope is an obvious next step.

In turn, the $N$-body generalization forms a prerequisite for the development of the general theory with a varying number of bodies.

A major limitation of the theory here is the assumption that the relevant quantities—the (invariant) probability densities, timeshifts, vector fields, etc.—are smooth. Constructing the relativistic 1-body generalization in the category of smooth manifolds first allowed us to develop the conceptual structure without the need to worry about the regularity of the quantities in question. Yet ultimately an acceptable relativistic 1-body quantum theory will need to take a functional-analytic perspective with regards to the quantities $\rho$ and $V$ (respectively $\varrho$ and $\iota$), making use of appropriate Lebesgue and Sobolev spaces to allow for a sensible notion of ‘weak solutions’ (cf. Remarks 2.2.(4), and 3.5 below; see also Refs. [142,172]).

In this respect, we note again that this work remains mostly agnostic with regards to any specific dynamical models. On the one hand, it is a strength, for the theory provides a language in terms of which a wide variety of dynamical models can be formulated as long as the latter comply with the basic assumptions.[35] On the other hand, it is a weakness, for physics fundamentally concerns itself with dynamics. Kinematics is only a means to an end. It remains to be seen, whether this work provides any insight with regards to the physical (in)acceptability of the Dirac equation as a 1-body theory for the electron beyond what is discussed below. In either case, it is certainly a natural starting point for addressing the much broader question of dynamics.

On a final note, we expect the questions of regularity and dynamics to be interrelated based on our experience with partial differential equations (see also Ref. [172]).

### 1.4 Further remarks on the literature

The earliest reference that we could locate in the literature discussing a non-trivial candidate for a relativistic Born rule is a 1934 article by Bloch [20]. On p. 314 sq. therein he notes that the many-time wave function, which Dirac, Fock, and Podolsky introduced in their 1932 article [42],[36] plays the role of a “probability amplitude.” The physical relevance of this ‘many-time formalism’ is, however, rather questionable—not only did the authors introduce one time for each body but also a “common time” and a “field time” for a single electromagnetic field supposed to act on all bodies. One may regard those works as historical attempts to resolve the ‘problem of time’ (discussed in Sect. 1.2 above).

Since the theory of the relativistic 1-body Born rule is an exemplar of the theory of the relativistic continuity equation, the theory of relativistic fluid mechanics constitutes a part of the literature on the subject. In this context, various authors have been studying if and how scalar quantities such as mass, entropy, and charge are conserved under temporal evolution.[37] Given that fluid dynamics is a standard topic in relativity theory, it should not come as a surprise that the most important prior mathematical contributions [55,59] to the relativistic 1-body Born rule have been made in this subject area.

[35] The second author of this work suggested a respective special-relativistic dynamical model in Ref. [165]. See also Refs. [166,204].

[36] Already in Ref. [45] Dirac had introduced the idea of using one time for each body to preserve relativistic invariance.

More direct treatments of the relativistic Born rule can be grouped into two main categories:

(1) Lienert, Lill, and Tumulka's treatment of the special-relativistic $N$-body case [132,133].
(2) In the more recent literature, Miller, Eckstein, and coauthors pursued a measure-theoretic approach to both the 1-body [56–58,142,143] and the $N$-body Born rule [144]. In Ref. [57], Eckstein and Miller set the respective mathematical foundations by generalizing the causal relations between points on a spacetime to measures thereon. This was later used to show the invariance theorems for the 1- and $N$-body case we mentioned in Sect. 1.2 above.

With regards to the $N$-body Born rule, the authors of the respective main references [132,144] ought to be credited for pursuing a 'single time parameter' approach. Despite the flaws of those constructions, the Lagrangian picture indeed makes such an approach feasible.

Returning to the broader literature, it is justified to say that the 1-body theory has been studied directly or indirectly in a variety of contexts and under many different assumptions.

Despite this diversity, there are two problematic assumptions that authors commonly make when treating the subject: That the spacetime ought to be globally hyperbolic[38] and that the respective hypersurface ought to be spacelike. The former assumption is generally accompanied by the latter.[39] This work demonstrates that those two assumptions are too restrictive. A more in-depth criticism is given in Remark 2.1 below.

Finally, we shall make some remarks on various "no-go theorems" that have appeared in the literature [93,94,138] and that one may view as relevant to this work.[40] Indeed, they have received some attention in the recent literature [39,68,132].

In Ref. [138], Malament argues that under a certain set of assumptions a relativistic generalization of the Born rule always has to yield the result zero—which would clearly defeat its purpose. The reader is invited to check that the respective result does not apply to the theory in this work. More generally, we view it as rather problematic from a methodological point of view to claim a general "theorem" that is beyond the reach of pure mathematics—one should not mistake one's own physical interpretation of a mathematical statement as the statement itself.[41]

Hegerfeld's work [93] is of more direct interest here, for he shows that in QM the infinite "propagation speed" of initial wave functions with compact support is essentially a result of the Hamiltonian formalism.[42] Accordingly, Hegerfeld's result might indicate a serious limitation of the Hamiltonian formalism. Since this formalism is intimately tied to the linearity of the 'general Schrödinger equation'

$$\mathrm{i}\hbar \frac{\partial \Psi}{\partial t} = \hat{H}\Psi, \tag{1.3}$$

[37] See e.g. Ref. [55], p. 25 sq., p. 38 sq., and p. 50 sq. in Ref. [59], p. 69 in Ref. [92], and Refs. [19,131,139,182,196,200].

[38] See e.g. Refs. [40,107,128,129,132,143,144].

[39] See Refs. [35,89,91,103,145,179,192,198,205] for treatments that require spacelikeness of the respective hypersurface without explicitly assuming that the spacetime is globally hyperbolic. Similarly, Ref. [19] requires non-degeneracy of the induced metric.

[40] One should think of the projection operators the authors refer to as multiplication with the characteristic function of a (measurable) subset of the respective hypersurface.

[41] To be fair, the author makes it clear that one does not need to agree with his view on the matter, so that our objection mainly applies to the title of the work. There are, however, multiple other works in the foundations of quantum theory that are misleading in this sense.

[42] See also Refs. [81,137].

in relativistic quantum theory one may need to consider non-linear evolution equations from the onset.[43,44] Since the subject of dynamics is not the focus of this work, however, we shall not pursue this matter any further here.

## 1.5 Notations and conventions

For the reader's convenience, we outline some notations and conventions.

With a few notable exceptions, we generally follow Ref. [177] in this respect. Unless otherwise noted, all mappings and manifolds in this work are assumed to be smooth.

We use the $(+ - \cdots -)$-convention for the metric and Einstein summation convention with lower case Latin indices. Though we have placed much emphasis on rigor, at times we do use 'sloppy language' that omits relevant structures—for instance, we might state that "$\mathcal{S}_0$ is a submanifold of the spacetime $\mathcal{Q}$" as opposed to "$(\mathcal{S}_0, \iota_0)$ is a submanifold of the spacetime $(\mathcal{Q}, g, \mathcal{O})$." We also do not distinguish between $\mathbb{R}$- or $\mathbb{C}$-valued functions on a manifold and respective scalar fields (i.e. global sections of the respective trivial bundle). If we use the word 'initial' in connection with a submanifold or hypersurface, this is made with a reference to its 'temporal evolution'; it does not refer to the mathematical property of being 'weakly embedded' (cf. p. 113 sq. in Ref. [127] or Def. 1.6.9 in Ref. [177]).

Due to the importance of integrals in this work, we point out that our definition of the integral coincides with the one in Def. 4.2.6 in Ref. [177]. Moreover, at times we also employ the simplified notation for integrals over submanifolds stated in Def. 4.2.7 thereafter. Locally, $\int \mathrm{d}^n\kappa$ is a shorthand notation for the $n$-fold (Lebesgue) integral with respect to the coordinates $\kappa$. For a manifold $\mathcal{Q}$ the set $\mathcal{B}^*(\mathcal{Q})$ denotes its Lebesgue $\sigma$-algebra (cf. Footnote 49 below).

Quite often, we make use of the term 'natural inclusion': If $B$ is a set and $A$ is a subset of $B$, then the *natural inclusion (of $A$ into $B$)* is the map $A \to B : q \mapsto q$. In the context of manifolds, this map is usually required to be smooth—though it need not be a topological embedding. Its local coordinate representative usually differs from the identity mapping.

With regards to general differential-geometric notation, we use $\mathrm{d}$ for the exterior derivative and $\mathcal{L}_Y$ for the Lie derivative along some vector field $Y$. Furthermore, if $\varphi$ is a smooth mapping (between manifolds), its pullback is $\varphi^*$, its pushforward is $\varphi_*$ and $\varphi|_A$ is the map with domain restricted to $A$. A generic notation for its domain is $\operatorname{dom}\varphi$ and for its graph it is $\operatorname{graph}(\varphi)$. At times we find it useful to use a period as a placeholder, e.g. $\varphi(\,.\,) = \varphi$. The dot $\cdot$ denotes matrix multiplication and we use $\lrcorner$ for the interior product. The transpose of a matrix $A$ is $A^{\mathrm{T}}$ and $\mathbb{1}$ is some kind of identity, as determined by the context. The tangent, cotangent and $k$-fold exterior algebra bundles of a manifold $\mathcal{Q}$ are denoted by $\mathrm{T}\mathcal{Q}$, $\mathrm{T}^*\mathcal{Q}$, and $\Lambda^k\,\mathrm{T}^*\mathcal{Q}$, respectively. Anti-symmetrization of tensor components is denoted by square brackets; if for instance $\alpha_{i_1 i_2}$ denotes the components of a 2-form, then $\alpha_{[i_1 i_2]} \equiv \alpha_{i_1 i_2}$. We also employ the usual notation for the Kronecker delta and the Levi-Civita symbol.

Finally, we note that $c$ denotes the speed of light (in vacuum).

[43] If one views the Madelung equations as foundational to the non-relativistic theory [169,172], then a non-linear approach is indeed quite natural. In particular, the results of Sec. 6 in Ref. [169] suggest that a variation of the number of bodies over time necessarily forces the theory to be non-linear.
We refer to Sec. 4 in Ref. [169] as well as Secs. 2.3 and 2.4 in Ref. [170] for a discussion of how observables may be treated in such a theory.

[44] Halvorson and Clifton [90] credit Barrett [11] for suggesting to "abandon unitary dynamics" in order to address Malament's [138] and Hegerfeld's objections [93]. The authors dismiss Barrett's suggestion as "little more than wishful thinking" based on the assertion that such a theory cannot be expected to reproduce "quantum interference effects."
It should not be controversial to say that the linear quantum-mechanical evolution equations need to be re-obtained from corresponding relativistic equations in the Newtonian limit—independent of whether the relativistic equations themselves are linear or not. As Landé has elaborated on in Refs. [125,126], this is sufficient to yield the "quantum interference effects" that Halvorson and Clifton [90] insist on.

## 2 The instantaneous 1-body Born rule

Due to subtle differences in the use of terminology in the mathematical general relativity literature, we shall begin with a few definitions. Readers looking for an introduction to the general theory of relativity are directed to Refs. [92,148,153,179,208]. General introductions to the relevant differential geometry may be found in Refs. [127,176,177].

The first term we consider is that of a spacetime, arguably the central concept of relativity theory. Since it is primarily a physical concept, different mathematical definitions may be found in the literature. Notwithstanding questions of regularity, the definition provided here is intended to be as wide as possible while including mathematical structures that are essential for the physics, regardless of the particular model in question.

**Definition 2.1** A *spacetime* is a tuple $(\mathcal{Q}, g, \mathcal{O})$ such that $(\mathcal{Q}, g)$ is a Lorentzian manifold and $\mathcal{O}$ is a spacetime orientation that is compatible with $g$ (as explained below). Moreover, if $\mathcal{Q}$ has dimension $(n+1)$ with $n \in \mathbb{N}$, we call the respective spacetime an $(n+1)$-*spacetime*. ◇

As we do not intend to go into the theory of principal bundles here, one may indirectly define a compatible spacetime orientation $\mathcal{O}$ by declaring some global timelike vector field to be 'future-directed' and fixing an 'ordinary' orientation on $\mathcal{Q}$.[45] Thus, only orientable manifolds that admit a global nowhere-vanishing vector field can be turned into spacetimes. The physical justification for the above definition is that it allows one to distinguish both 'past' from 'future' and 'right-handed' from 'left-handed'.

The second term we shall define already constitutes a first step towards formulating the general-relativistic Born rule: By viewing time in Newtonian mechanics as a separate coordinate rather than a parameter, one finds that for the sought-for generalization one requires an $n$-dimensional submanifold of an $(n+1)$-spacetime which in some sense represents 'space'—roughly speaking, a 'hypersurface' subject to further conditions.

**Definition 2.2** Let $\mathcal{Q}$ be a manifold of dimension at least 1. Then, a *hypersurface* of $\mathcal{Q}$ is an embedded submanifold $(\mathcal{S}_0, \iota_0)$ of $\mathcal{Q}$ of dimension $\dim \mathcal{Q} - 1$. ◇

The noteworthy requirement is that $\iota_0$ ought to be a topological embedding. Apart from providing some important technical advantages,[46] the embeddedness property allows one to think of the submanifold as a subset of $\mathcal{Q}$. This is important because we are interested in the probability that the body is located in a 'region' of the image $\iota_0(\mathcal{S}_0)$ in $\mathcal{Q}$; the manifold $\mathcal{S}_0$ by itself is only an auxiliary object.

Still, it is not always convenient to view $\mathcal{S}_0$ as a subset of $\mathcal{Q}$. For instance, as a model for temporal evolution of the hypersurface $(\mathcal{S}_0, \iota_0)$, the kinematic case will supply us with examples in which the manifold $\mathcal{S}_0$ embeds into $\mathcal{Q}$ via mappings that differ from the natural inclusion (even if $\mathcal{S}_0 = \iota_0(\mathcal{S}_0)$).

Having defined these two important terms, we shall proceed to generalize the Born rule for the 'instantaneous case', i.e. the case for which we do not account for the temporal evolution of the system.

As explained, this Born rule has to be formulated on a hypersurface $(\mathcal{S}_0, \iota_0)$ in an $(n+1)$-spacetime $(\mathcal{Q}, g, \mathcal{O})$. As in the non-relativistic theory, it will be stated as an integral over 'measurable' subsets of $\mathcal{S}_0$. Therefore, the mathematical task is to construct a suitable integrand and guarantee that the respective integral is well-defined.

To do so, it is of use to recall the general principle of relativity stated in Sect. 1.2 above. Taken literally, adherence to it can be assured by employing the language of Cartan calculus to construct the central mathematical objects of the theory from more basic ones, the latter of which may also not rely on a particular choice of coordinates.

We, therefore, require a coordinate-independent quantity that mathematically encodes the position probability on the hypersurface. Inspired by the relativistic theory of electromagnetism and the Dirac equation, we take this quantity to be the (probability) current density (vector) field $J$. In the instantaneous case, we only require its values $J_0$ on the hypersurface $\iota_0(\mathcal{S}_0)$, however. That is, $J_0$ is assumed to be a vector field over $\mathcal{S}_0$, i.e. it is a (smooth) section of the pullback bundle

$$\iota_0^* \mathrm{T}\mathcal{Q} = \left\{(q, Y) \in \mathcal{S}_0 \times \mathrm{T}\mathcal{Q} \middle| Y \in \mathrm{T}_{\iota_0(q)}\mathcal{Q}\right\} \tag{2.1}$$

[45] Generally, this does not ascribe any physical significance to the vector field that goes beyond the aforementioned utility. For a full definition of spacetime orientations we refer to Sec. 2.2 in Ref. [167].

[46] First and most importantly, embeddedness is a necessary assumption for the applicability of the flowout theorem (Thm. 9.20 in Ref. [127]). The importance of the latter to the general theory will be shown in Sect. 3 below. Second, it conveniently guarantees the existence of slice charts (cf. p. 101 sqq. in Ref. [127]).

(cf. Sec. 2.6 in Ref. [177]).

Yet for $J_0$ to be a physical current density field over $\mathcal{S}_0$, we must require that at each point $q \in \mathcal{S}_0$ the vector $(J_0)_q$ either vanishes or is future-directed causal. The underlying physical reason is that $J_0$ determines the 'propagation direction' of probability and this propagation ought not to be "faster than the speed of light $c$" [207] (i.e. it ought not to violate the physical condition of causality).

The 'length' of $J_0$ also has a physical interpretation: Given a 'measurable' subset $U$ of $\mathcal{S}_0$, then the 'length' of $J_0$ on $U$ determines the amount of probability contained in $U$ (cf. Sect. 3). Therefore, the set of points $q$ in $\mathcal{S}_0$ for which $(J_0)_q$ vanishes ought not to contribute to the integral.

By choosing $\mathcal{S}_0$ appropriately, we may therefore assume that $J_0$ is future-directed causal everywhere on $\mathcal{S}_0$.

Given such a $J_0$ and the canonical volume form $\mu$ on $\mathcal{Q}$ (as induced by $g$ and $\mathcal{O}$), we may then define the $n$-form[47]

$$\frac{1}{c}\,\iota_0^*\big(J_0 \lrcorner\, \mu_{\iota_0(\,.\,)}\big) \tag{2.2}$$

on $\mathcal{S}_0$. Clearly, this integrand does not depend on any particular choice of coordinates and it also does not depend on the particular choice of $(\mathcal{S}_0, \iota_0)$.

To our knowledge, the first person to suggest this integrand in the published literature was Ehlers (cf. p. 25 sq. and p. 50 sq. in Ref. [59]). Though he contributed to the theory of relativistic fluid mechanics rather than the theory of the relativistic Born rule directly, the close relationship between the former and the 1-body case of the latter justifies this credit (cf. Sect. 1.3).

However, if one were to use the integrand from Eq. (2.2) without imposing any further conditions on $\mathcal{S}_0$, it would be possible to choose $\mathcal{S}_0$ to be tangent to $J_0$—in which case the integrand can be shown to vanish everywhere. This, in turn, prevents the definition of a probability measure on $\mathcal{S}_0$ via the use of said integrand.

Contrarily and as we shall show below, if $J_0$ is nowhere-tangent to $\mathcal{S}_0$, then we may indeed normalize the respective integral. In particular, the mere existence of such a vector field on $\mathcal{S}_0$ assures that $\mathcal{S}_0$ is orientable (cf. Thm. 15.21 in Ref. [127]), which is a necessary condition for the integral to be mathematically defined.

We call this requirement the *transversality condition*.

One may view the transversality condition as the mathematical expression of our initial assertion that $(\mathcal{S}_0, \iota_0)$ ought to represent 'space' in some sense. The example in Fig. 2 illustrates that the condition indeed excludes certain hypersurfaces that do not meet the latter intuitive requirement.

*Remark 2.1* We shall consider the particular case that $(\mathcal{S}_0, \iota_0)$ is spacelike and orientable.

In that instance, the metric $g$ on $\mathcal{Q}$ induces a Riemannian metric $\tilde{g} = -\iota_0^* g$ on $\mathcal{S}_0$, which in turn gives rise to a volume form $\tilde{\mu}$ on $\mathcal{S}_0$. Denoting the unique future-directed timelike normal vector field over $\mathcal{S}_0$ by $\tilde{n}$, we may rewrite the integrand in Eq. (2.2) to[48]

$$\frac{1}{c}\, g_{\iota_0(\,.\,)}(J_0, \tilde{n})\ \tilde{\mu}. \tag{2.3}$$

[47] For every $q \in \mathcal{S}_0$ one finds that $(J_0)_q \lrcorner\, \mu_{\iota_0(q)}$ is an element of $\bigwedge^n \mathrm{T}^*_{\iota_0(q)}\mathcal{Q}$. Upon viewing $\iota_0^*$ evaluated at $q$ as a map from $\bigwedge^n \mathrm{T}^*_{\iota_0(q)}\mathcal{Q}$ to $\bigwedge^n \mathrm{T}^*_q\mathcal{S}_0$, one deduces that the $n$-form in Eq. (2.2) is indeed well-defined. Moreover, it is smooth by virtue of being a composition of smooth maps. See Eq. (A.1) below for an explicit expression.

[48] The proof of this identity is analogous to the Riemannian case, see e.g. Lem. 16.30 in Ref. [127].

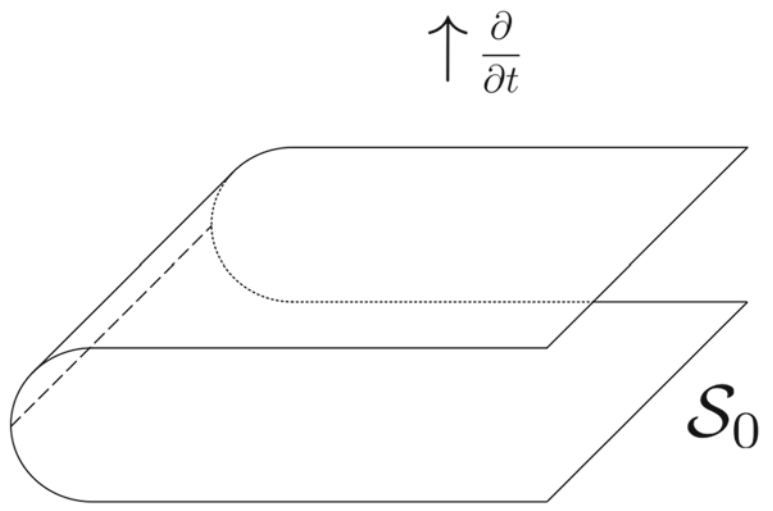

**Fig. 2** The sketch shows a hypersurface $\mathcal{S}_0$ in $(2+1)$-dimensional Minkowski spacetime. At the longly dashed line $\mathcal{S}_0$ is tangent to $\partial/\partial t$, so $\partial/\partial t$ cannot be used to define an orientation on $\mathcal{S}_0$. In fact, as any Euclidean product of any future-directed causal vector field $J_0$ with the respective Euclidean normal vector field on $\mathcal{S}_0$ has to change sign and this is continuous by assumption, the transversality condition cannot be satisfied on $\mathcal{S}_0$—independent of the particular choice of $J_0$

This is indeed the integrand commonly found in the literature.

The integrands in Eqs. (2.2) and (2.3) are, however, not equivalent: If one drops the condition that $(\mathcal{S}_0, \iota_0)$ is spacelike, then there may exist points $q$ in $\mathcal{S}_0$ for which $(\iota_0)_* \, \mathrm{T}_q \mathcal{S}_0$ is a lightlike or even timelike subspace of $\mathrm{T}_{\iota(q)} \mathcal{Q}$. In the former case the pullback metric is degenerate at $q$ and thus does not give give rise to a volume form. In the latter case, there are two normal vectors $\tilde{n}_q$, both of them spacelike.

Indeed, there are a number of conceptual problems resulting out of the requirement that $(\mathcal{S}_0, \iota_0)$ must be spacelike:

(1) First, it implicitly and fallaciously associates spacelike hypersurfaces in relativity theory with the Newtonian conception of space. This is problematic, because in non-relativistic physics $c$ is effectively infinite and thus the distinction between lightlike and spacelike hypersurfaces becomes meaningless therein. In a prior work, the first author of this work has argued that 'Newtonian space' is more appropriately associated with the past lightcone of the (pointlike) physical observer at an event. We refer to Sec. 3 in Ref. [167].

(2) The second problem is one of existence and also concerns the assumption of global hyperbolicity on the spacetime (cf. Sect. 1.4):
In 1965 Penrose [162] discovered the following property of plane-wave spacetimes:

No spacelike hypersurface exists in the space-time which is adequate for the global specification of Cauchy data.

We also refer to Ref. [163] and Chap. 13 in Ref. [12] for a discussion.
While the argument that such spacetimes are strong idealizations of actual gravitational radiation does have merit, it nonetheless constitutes a weak, physical argument for restricting oneself to spacelike hypersurfaces in globally hyperbolic spacetimes. For the purposes of formulating a quantum theory on curved spacetime, what is the physical argument that in plane-wave spacetimes there should not be any hypersurface "which is adequate for the global specification of Cauchy data" [162]?
It seems difficult to justify the two aforementioned assumptions, given that the mathematical theory outlined here exposes them as overly restrictive.

(3) The third problem is that in general spacelike hypersurfaces need not stay spacelike under the flow of a timelike/causal/lightlike vector field. We refer to Example 3.1 below.

The misconception explained in point (1) above seems to be the main reason why the relativistic 1-body Born rule had not been stated in its full generality in the literature before. In this respect, it is worth pointing out that Ehlers did not require the transversality condition when suggesting the integrand (2.2), though he might have merely omitted it (cf. p. 25 sq. and p. 50 sq. in Ref. [59]). ◇

Having gathered the main ingredients for the 1-body Born rule in the instantaneous case, we are ready to formulate the central theorem of this section.

**Theorem 2.1** *Let $(\mathcal{Q}, g, \mathcal{O})$ be a spacetime. Further, let $(\mathcal{S}_0, \iota_0)$ be a hypersurface in $\mathcal{Q}$ and let $J_0$ be a nowhere tangent, future-directed timelike/causal/lightlike vector field over $\mathcal{S}_0$.*

*The following holds:*

(1) $\mathcal{S}_0$ *is orientable and carries a canonical orientation.*
(2) *The expression in Eq.* (2.2) *is a volume form on* $\mathcal{S}_0$.
(3) *Let* $J_0$ *satisfy*

$$\frac{1}{c}\int_{\mathcal{S}_0} \iota_0^*\big(J_0 \lrcorner\, \mu_{\iota_0(\,.\,)}\big) = 1, \tag{2.4a}$$

*and denote by* $\mathcal{B}^*(\iota_0(\mathcal{S}_0))$ *the* $\sigma$*-algebra of Lebesgue subsets*[49] *of the image* $\iota_0(\mathcal{S}_0)$*. Then*

$$\mathbb{P}_0 \colon \mathcal{B}^*(\iota_0(\mathcal{S}_0)) \to [0,1] \colon U \mapsto \mathbb{P}_0(U) = \frac{1}{c}\int_{\iota_0^{-1}(U)} \iota_0^*\big(J_0 \lrcorner\, \mu_{\iota_0(\,.\,)}\big) \tag{2.4b}$$

*defines a probability measure on the measurable space* $\big(\iota_0(\mathcal{S}_0), \mathcal{B}^*(\iota_0(\mathcal{S}_0))\big)$*. Moreover, for* $U \in \mathcal{B}^*(\iota_0(\mathcal{S}_0))$ *the probability* $\mathbb{P}_0(U)$ *is* 0 *if and only if* $U$ *is a Lebesgue null set.* ◇

The first statement in point (3) of Theorem 2.1 states that $\mathbb{P}_0(U)$ is indeed a probability in the mathematical sense of the word. Accordingly, for the instantaneous case the (general-relativistic) 1-body Born rule states that $\mathbb{P}_0(U)$ is the probability that the body is located in $U$. Points (1), (2), and the second statement in point (3) are primarily of technical importance.

It is worth pointing out that there is no time given in the prescription $\mathbb{P}_0(U)$. The reason is that 'the time' is implicit in the choice of $\mathcal{S}_0$—which, in accordance with the general principle of relativity, is largely arbitrary.

Formally, we still need to prove that Theorem 2.1 relies on mathematically sensible assumptions.

**Proposition 2.1** *Let* $(\mathcal{Q}, g, \mathcal{O})$ *be a spacetime.*

(1) *There exists an orientable hypersurface* $(\mathcal{S}_0, \iota_0)$ *and a future-directed timelike/causal/lightlike vector field* $J_0$ *over* $\mathcal{S}_0$ *such that* $J_0$ *is nowhere tangent to* $\mathcal{S}_0$.
(2) *If* $J$ *is a future-directed timelike/causal/lightlike vector field on* $\mathcal{Q}$*, then there exists an orientable hypersurface* $(\mathcal{S}_0, \iota_0)$ *such that* $J_0 = J_{\iota_0(\,.\,)}$ *is nowhere tangent to* $(\mathcal{S}_0, \iota_0)$.

◇

In general, for a given orientable hypersurface $(\mathcal{S}_0, \iota_0)$ in $\mathcal{Q}$, there need not exist any nowhere-tangent causal vector field over $\mathcal{S}_0$. Figure 2 again provides an example.

For the reader's convenience, we give some coordinate expressions.

**Lemma 2.1** *Consider the situation of Theorem* 2.1 *with* dim $\mathcal{Q} = n + 1$ *for* $n \in \mathbb{N}$.

(1) *Let* $(U, \kappa)$ *be an oriented slice chart for* $\mathcal{S}_0$ *in* $\mathcal{Q}$*. Denote by* $\vec{\kappa} \mapsto (0, \vec{\kappa})$ *the respective local coordinate expression of* $\iota_0$*. Then for all* $W \in \mathcal{B}^*(\iota_0(\mathcal{S}_0) \cap U)$ *we have*

$$\mathbb{P}_0(W) = \frac{1}{c}\int_{\vec{\kappa}(W)} J_0^0(\vec{\kappa})\, \sqrt{-\det g}\big(0, \vec{\kappa}\big)\, \mathrm{d}^n\kappa. \tag{2.5a}$$

(2) *Let* $(V, \xi)$ *and* $(U, \kappa)$ *be oriented charts on* $\mathcal{S}_0$ *and* $\mathcal{Q}$*, respectively, such that* $\iota_0(V) \cap U$ *is nonempty. Denote by* $\xi \mapsto \kappa(\xi)$ *the respective local coordinate expression of* $\iota_0$*. Then, for all* $W \in \mathcal{B}^*\big(\iota_0(V) \cap U\big)$ *we have*

[49] We refer to Sec. XII.1 Ref. [3] for an definition of and elaboration on the Lebesgue $\sigma$-algebra on a manifold and its Lebesgue null sets.

$$\mathbb{P}_0(W) = \frac{1}{c}\int_{\xi\left(\iota_0^{-1}(W)\right)} (n+1)!\, J_0^{[0}(\xi)\,\frac{\partial\kappa^1}{\partial\xi^1}(\xi)\cdots\frac{\partial\kappa^{n]}}{\partial\xi^n}(\xi)\,\sqrt{-\det g\big(\kappa(\xi)\big)}\,\mathrm{d}^n\xi\,. \tag{2.5b}$$

◇

The following example is meant to illustrate how the relativistic Born rule is to be used in practice and that it is indeed of relevance to relativistic quantum theory.

*Example 2.1* Consider Minkowski 4-spacetime $\left(\mathbb{R}^4, g, \mathcal{O}\right)$ with standard coordinates $x = (ct, \vec{x})$. As noted above, one may indirectly define the spacetime orientation $\mathcal{O}$ by declaring that the standard coordinates are oriented and that the timelike vector field $\partial_0 = \partial/\partial(ct)$ is future-directed.[50]

For $i \in \{0, 1, 2, 3\}$ denote by $\gamma^i$ the $i$th 'gamma matrix'. Let $\Psi$ be a section of the trivial bundle $\mathbb{R}^4 \times \mathbb{C}^4$. We think of $\Psi$ as a (smooth) solution of the Dirac equation, possibly in the presence of an external electromagnetic field. We refer, for instance, to Ref. [97] or Sec. 12.2 in Ref. [99].

As indicated in Sect. 1.1, there is some discussion in the physics literature on the interpretation of such 'Dirac spinor fields' $\Psi$ and whether the Dirac equation is an acceptable 1-body evolution equation or not (see e.g. p. 1 and Chap. 1 in Ref. [201]). While this question is ultimately a physical one, Holland [97] has made a convincing argument why (sufficiently regular) Dirac spinor fields do indeed give rise to a mathematically acceptable 1-body description:

The first point is that the Dirac current $J$ with components

$$J^i = \Psi^\dagger \cdot \gamma^0 \cdot \gamma^i \cdot \Psi \tag{2.6a}$$

yields a future-directed timelike vector $J_x$ at any $x \in \mathbb{R}^4$ for which $\Psi(x) \neq 0$. Thus, if we define the open set

$$\mathcal{Q} = \mathbb{R}^4 \setminus \Psi^{-1}(\{0\}) \tag{2.6b}$$

and restrict $g$, $\mathcal{O}$, $\Psi$, and $J$ accordingly, then $J$ becomes a future-directed timelike vector field on the spacetime $\mathcal{Q}$.

The second such point that speaks for the Dirac theory is that $J$ satisfies the relativistic 1-body continuity equation and that thus the total probability is indeed conserved. A more detailed discussion of this point may be found in Sect. 3.3 below; the particular result is given in Corollary 3.1.(4).

These two points make it mathematically possible and sensible to discuss the relativistic 1-body Born rule within the Dirac theory. In order to keep the discussion here simple, we shall assume, as above, that $\Psi$ and hence $J$ do not vanish.[51] Still, Eq. (2.6i) below holds true regardless.

Having determined $J$ from such a solution $\Psi$ of the Dirac equation, one needs to assure that $J$ is normalized—as one normalizes wave functions in QM. In practice, this is done on the given initial hypersurface $(\mathcal{S}_0, \iota_0)$ and then probability conservation guarantees normalization for all other 'times'. Entirely out of convenience we use

$$\frac{1}{c}\int_{\mathbb{R}^3} J^0(0, \vec{x})\,\mathrm{d}^3x = 1 \tag{2.6c}$$

(cf. Lemma 2.1.(1)).

Assume now that a (pointlike) physical observer is positioned at $x_0 = (ct_0, \vec{x}_0) \in \mathbb{R}^4$ and is able to detect the particle in its past light cone $\mathcal{S}_0$. Physically, such a detection can only be realized by the interaction of the particle with other matter or radiation, but for the sake of explaining the mathematical theory we shall ignore this complication (see also Footnote 14 above).

---

50 $\mathcal{O}$ is a reduction of the (trivial) frame bundle of $\mathbb{R}^4$ and is chosen such that the coordinate frame field $\partial = (\partial_0, \ldots, \partial_3)$ with respect to $x$ is a global section thereof.

51 The discussion also applies, if $J$ only vanishes on a set of measure zero on an admissible hypersurface.

Accordingly, we consider the hypersurface[52]

$$\mathcal{S}_0 = \left\{ (ct, \vec{x}) \in \mathbb{R}^4 \middle| t = t_0 - |\vec{x} - \vec{x}_0| / c < t_0 \right\} \tag{2.6d}$$

together with the natural inclusion $\iota_0$. It may be shown that the coordinate functions $\vec{x}$ can be restricted to yield coordinates on $\mathcal{S}_0$, so that with respect to these coordinates, we have

$$\iota_0(\vec{x}) = (ct_0 - |\vec{x} - \vec{x}_0|, \vec{x}) = (x^0(\vec{x}), \vec{x}) \tag{2.6e}$$

for all $\vec{x} \in \mathbb{R}^3 \setminus \{0\}$.

As we are only interested in the values of $J$ over $\mathcal{S}_0$, the quantity $J$ in the integral in Eq. (2.5b) will be replaced by $J_0 = J \circ \iota_0$. The coordinate representation of the latter is given by

$$J_0 \colon \vec{x} \mapsto (J_0)_{\vec{x}} = \left( J_0^0(\vec{x}), \vec{J}_0(\vec{x}) \right) = \left( J^0(\iota_0(\vec{x})), \vec{J}(\iota_0(\vec{x})) \right). \tag{2.6f}$$

To compute the respective integrand, we observe that

$$\begin{aligned} &(3+1)!\, J^{[0}(\iota_0(\vec{x})) \frac{\partial x^1}{\partial x^1}(\vec{x}) \frac{\partial x^2}{\partial x^2}(\vec{x}) \frac{\partial x^{3]}}{\partial x^3}(\vec{x}) \\ &= \det \begin{pmatrix} J^0(\iota_0(\vec{x})) & -(\vec{x} - \vec{x}_0)^{\mathrm{T}} / |\vec{x} - \vec{x}_0| \\ \vec{J}(\iota_0(\vec{x})) & \mathbb{1} \end{pmatrix}. \end{aligned} \tag{2.6g}$$

For a Lebesgue set $U' \subseteq \mathbb{R}^3$, the probability that the body is located in the region

$$U = \iota_0(U') = \left\{ (ct, \vec{x}) \in \mathbb{R}^4 \middle| t = t_0 - |\vec{x} - \vec{x}_0| / c < t_0 \text{ and } \vec{x} \in U' \right\} \tag{2.6h}$$

of the past light cone of the observer at $x_0$ is, therefore,

$$\mathbb{P}_0(U) = \frac{1}{c} \int_{U'} \left( J^0(\iota_0(\vec{x})) + \frac{\vec{J}(\iota_0(\vec{x})) \cdot (\vec{x} - \vec{x}_0)}{|\vec{x} - \vec{x}_0|} \right) \mathrm{d}^3 x. \tag{2.6i}$$

In practice it may, however, be simpler to use Corollary 3.1.(5) below to compute the respective probability. ◇

We conclude the discussion of the instantaneous case with a few remarks.

*Remark 2.2* (1) Physical arguments have to be employed to show that the integrand in Eq. (2.2) is indeed the physically correct choice.
Identity (3.17c), which is discussed in Sect. 3.3, may be used as an argument for the 'naturalness' of this choice. The identity states that the integrand does not change along the direction of the flow of probability whenever the relativistic continuity equation, div $J = 0$, is satisfied.
Regardless, a necessary criterion for physical consistency is that the Newtonian limit of the relativistic theory yields the non-relativistic theory: In this limit, the non-relativistic 1-body Born rule must be reobtained, and the demand for probability conservation in the relativistic theory has to carry over to the non-relativistic one. Since the question of the rigorous Newtonian limit is beyond the scope of this article, we shall not discuss this question here. We refer to Sec. 4.2 in Ref. [167] for a rigorous approach.

[52] It is worth pointing out that the acceptability of $(\mathcal{S}_0, \iota_0)$ as a hypersurface for formulating the Born rule relies on Corollary 3.1.(4)—despite the fact that there is one point $x_0$ which is not intersected by a respective integral curve. This is not a problem, because for the purpose of integration, that point constitutes a set of zero measure. To be entirely rigorous, one may exclude the image of the respective integral curve from the spacetime.

(2) The transversality condition alone is in general not sufficient to exclude all hypersurfaces one would want to exclude physically.
For instance, in Example 2.1 the hypersurface

$$\mathcal{S}_1 = \left\{(ct, \vec{x}) \in \mathbb{R}^4 \middle| t = 0 \text{ or } t = t_1\right\} \tag{2.7}$$

with $t_1 \neq 0$ is clearly an unphysical choice, yet so far, we have not provided any mathematical condition to exclude it.
We will provide such a condition in Remark 3.4 below, noting here only that it is somewhat artificial to consider the Born rule in the absence of any temporal evolution, and that the justification for doing so is primarily pedagogical. We refer to Lemma 3.3 below for an elaboration on how the kinematic Born rule relates to the instantaneous one.
A related point is that one should also require $(\mathcal{S}_0, \iota_0)$ to be 'maximal', so that one normalizes $J$ over 'the entire space'. Under the additional assumption that $\mathcal{S}_0$ is connected, one may define $(\mathcal{S}_0, \iota_0)$ to be *maximal*, if there does not exist another connected hypersurface $(\mathcal{S}'_0, \iota'_0)$ such that $\iota_0(\mathcal{S}_0)$ is properly contained in $\iota'_0(\mathcal{S}'_0)$.[53]
Usually, however, an appropriate choice of 'initial hypersurface' is clear from as well as dictated by physical considerations. In that case, a rigorous mathematical definition would be more cumbersome than useful.

(3) For the purpose of measurement, one may want to impose more restrictions on the initial hypersurface $(\mathcal{S}_0, \iota_0)$: For instance, if $J$ is known to be future-directed timelike (which ought to be the case for particles with non-zero mass), yet the values of $J$ are a priori unknown, then the transversality condition can only be guaranteed by requiring $\mathcal{S}_0$ to be nowhere timelike.[54] This point does, however, not invalidate Remark 2.1 above.

(4) In the related literature some authors consider the Borel $\sigma$-algebra instead of the Lebesgue $\sigma$-algebra for formulating the Born rule on an appropriate manifold $\mathcal{S}_0$. The problem with this choice is that it does not yield a complete measure space, which is in turn required to sensibly define $L^p$-spaces on $\mathcal{S}_0$. $L^p$-spaces are important for the mathematical analysis of potential candidates for dynamical equations and, if sensible, for defining spaces of 'wave functions'.

◇

## 3 The kinematic 1-body Born rule

The kinematic Born rule generalizes the instantaneous one so as to allow for the temporal evolution of the initial hypersurface. The discussion in Sect. 3.1 below is intended to be a heuristic introduction to the core ideas and, with the exceptions of Proposition 3.1 and Example 3.1 below is, therefore, not rigorous. The respective mathematical theory will be developed in Sects. 3.2 and 3.3.

### 3.1 Conceptual approach

We begin by treating the question of how one would generally define the 'time evolution' of a hypersurface $(\mathcal{S}_0, \iota_0)$ in a spacetime $\mathcal{Q}$.

The reader may recall that in relativity theory the temporal evolution of a single point representing a point mass is generally modeled via a future-directed timelike curve—which is an observer curve in the particular case of proper time parametrization. Moreover, in the geometric optics approximation the propagation of electromagnetic radiation is usually modeled by (a family of) future-directed lightlike curves (cf. Sec. 3.2 in Ref. [184]). Thus, for the purpose of modeling the time evolution of a single point in a spacetime $\mathcal{Q}$ we should consider curves that are future-directed timelike, future-directed lightlike, and perhaps even future-directed causal.

[53] One may indeed exclude the example in Eq. (2.7) by requiring the initial hypersurface to be connected. Yet in conjunction with the requirement that $J$ is nowhere-vanishing this is a rather restrictive assumption. Ultimately, such problems point towards the need to go beyond the smooth theory, so that sets of measure zero in $\mathcal{S}_0$ become irrelevant.

[54] Note that a hypersurface that is timelike at a point is timelike in an open neighborhood thereof.
To see this, choose a timelike tangent vector on the hypersurface and smoothly extend it thereon to a vector field $Y$ (cf. Lem. 8.6 in Ref. [127]). Then the set $(g(Y, Y))^{-1}(\mathbb{R}_+)$ is such a neighborhood.

If we view the hypersurface as a collection of points, we may therefore model its evolution by asking that for each $q$ in $\mathcal{S}_0$ we have a real open interval $\mathcal{I}_q$ and a map $\iota$ such that

$$\iota(\,.\,, q)\colon \mathcal{I}_q \to \mathcal{Q}\colon \tau \mapsto \iota(\tau, q) \tag{3.1}$$

is a future-directed timelike, causal, or lightlike curve.

Moreover, we ask that the tuple $\big(\mathcal{S}_0, \iota(\tau,\,.\,)\big)$ is a hypersurface in its own right and that the curves do not intersect each other for different values of $\tau$. We also do not allow tearing, since, heuristically speaking, at the given 'parameter time' $\tau$ this would make $\iota$ discontinuous in $q$. Leaving the precise mathematical construction for Sect. 3.2 below, we find that the domain $\mathcal{S}$ of $\iota$ ought to be an open subset of $\mathbb{R} \times \mathcal{S}_0$, obtained by taking the disjoint union of all the intervals $\mathcal{I}_q$, and that $\iota$ itself ought to be a diffeomorphism to its image (or at least a local diffeomorphism; see Remark 3.1.(2) below).

Borrowing terminology from the non-relativistic continuum mechanics, this is indeed the general idea of how the temporal evolution is modeled in the *Lagrangian picture*. As in the non-relativistic analogue, the map $\iota$ fixes the initial point $q$ and then follows its trajectory as $\tau$ increases.[55]

In the non-relativistic theory, one may, however, take another point of view known as the *Eulerian picture*. In the relativistic theory, this is also a valid approach: Here the evolution of the hypersurface $(\mathcal{S}_0, \iota_0)$ is indirectly determined by a future-directed timelike, lightlike, or even causal vector field $V$ on $\mathcal{Q}$. The velocity vector field $V$ then gives rise to a flow $\Phi$,[56] which in turn connects the Eulerian picture with the Lagrangian picture via the relation

$$\iota(\tau, q) = \Phi_\tau\big(\iota_0(q)\big). \tag{3.2}$$

A schematic illustration of how the time evolution of a hypersurface may be modeled in the relativistic theory is given in Fig. 3.

Of course, the described approach is of little use, if it cannot be shown to be consistent with the instantaneous Born rule of Sect. 2. To make the two consistent, two points require elaboration: First, what is the relation between the vector fields $J$ and $V$? Second, does the suggested time evolution model respect the transversality condition of the hypersurface? After all and in accord with the general principle of relativity, the underlying theory should be independent of which hypersurface one chooses as initial.

Regarding the first point, recall that in the instantaneous case, we assumed that the dynamics of the respective 1-body quantum theory provides us with a future-directed timelike, causal, or lightlike current density vector field $J$ on $\mathcal{Q}$. It would, therefore, be mathematically sensible to set the two vector fields $J$ and $V$ equal.

Yet, this choice is physically incorrect, because the components of $J$ and $V$ have different physical dimensions: This can be seen by considering, for instance, the components $J^i$ of $J$ and $V^i$ of $V$, respectively, in standard coordinates in Minkowski spacetime. By assumption, each $J^i$ has the physical dimension of a probability current—velocity times inverse volume. The components $V^i$, on the other hand, appear in the integral curve equation, and thus have the physical dimension of velocity.

[55] Note that we do not necessarily suggest that subatomic particles follow these trajectories, as it is the case in the de Broglie-Bohm theory [22,23,49,50,99,181]. Rather the trajectories describe the evolution of the hypersurface itself and thus, by means of the continuity equation, the evolution of the probability density (cf. Refs. [62,169]). We also refer to Sec. 9.5 in in Ref. [21] as well as Ref. [38] for related discussions of interpretation.

[56] Complete timelike vector fields were studied by Dirmeier [47].

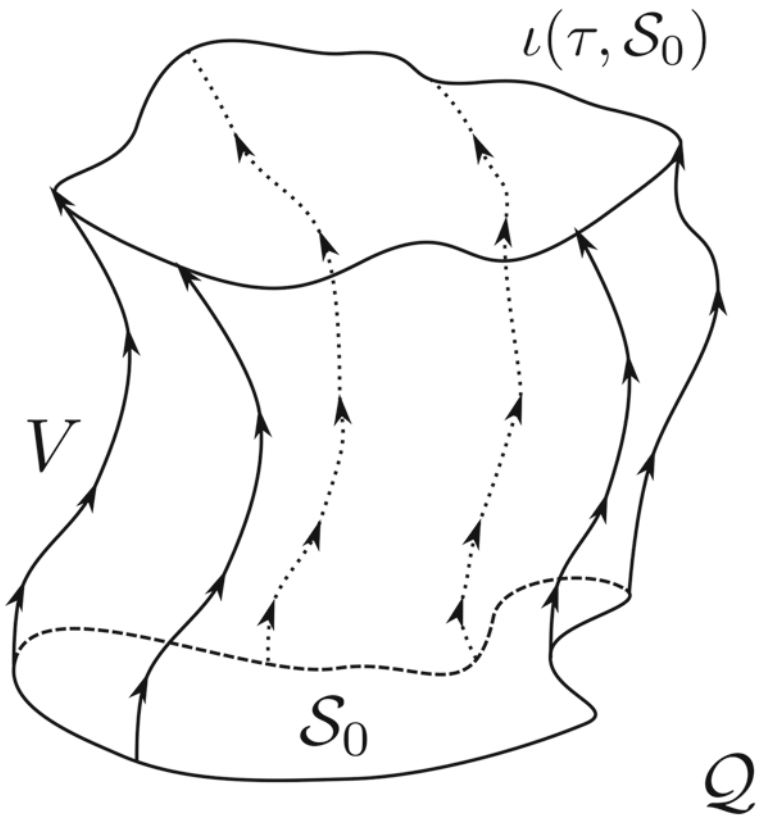

**Fig. 3** The sketch illustrates the general idea of how to define the time evolution of an initial hypersurface $\mathcal{S}_0$, here given as a subset of a spacetime $\mathcal{Q}$: In the Lagrangian picture, one evolves every point in $\mathcal{S}_0$ along a future-directed timelike, causal, or lightlike curve, as given by Eq. (3.1). Each 'parameter time' $\tau$ then gives rise to a new hypersurface, namely the image $\iota(\tau, \mathcal{S}_0)$. In the Eulerian picture, on the other hand, one considers the velocity field $V$ instead, which is a future-directed timelike, causal, or lightlike vector field on $\mathcal{Q}$ that is nowhere tangent to $\mathcal{S}_0$. The flow of $V$ relates the two pictures

Nonetheless, we do interpret the 'direction' of $J$ as the 'propagation direction' of probability. Therefore, we require that there exists a strictly positive—though usually not constant—factor of proportionality $\rho$:

$$J = \rho V. \tag{3.3}$$

Equation (3.3) is the relativistic analogue of the respective non-relativistic relation between a current density vector field on the one hand and the probability density and (drift) velocity vector field on the other—i.e. $\vec{j} = \rho\, \vec{v}$.

This, of course, raises the question of how to determine $\rho$ from $J$. In the analogue theories in which $J$ is a mass or a charge current density (cf. Remark 1.2), $J$ will be timelike and $\rho$ may then be obtained by requiring that the integral curves of $V$ are observer curves. That is, $V$ is future-directed and satisfies

$$g(V, V) = c^2. \tag{3.4}$$

In quantum theory, however, Eq. (3.4) is more difficult to justify. The 1-body Dirac theory discussed in Example 2.1, for instance, does not make any statements on the question.

In the absence of any other physical conditions, using Eq. (3.4) to determine $V$ from $J$ via Eq. (3.3) is arguably a natural choice.[57] Indeed, physically the condition is needed for obtaining the corresponding velocity vector field $\vec{v}$ in the non-relativistic limit. Ultimately, the theory we lay out in this work does not depend on this choice, however, with different choices amounting to a mere redefinition of the respective quantities (cf. Remark 3.6 below). In practical applications, one may therefore make a choice that is computationally convenient.

However one chooses the factor $\rho$, it will have the physical dimension of a probability density and, if one wants to bestow independent physical significance upon it, the choice should be independent of a particular coordinate system, again in accordance with the general principle of relativity. The first person to suggest the definition of such an invariant quantity was Eckart, who in the context of special-relativistic fluid mechanics introduced an invariant (inertial) mass density $\rho$ satisfying Eq. (3.3) for a mass current density $J$ by requiring $V$ to fulfill Eq. (3.4). Inspired by his work, in the context of relativistic quantum theory we suggest to call $\rho$ an *invariant probability density*, provided that the condition to obtain it from the probability current density $J$ is coordinate-invariant as well—as it is the case for Eq. (3.4).[58]

[57] This is particularly true if the dynamical equations are formulated in terms of $J$ only. Some results in the non-relativistic theory do indeed suggest this to be the physically correct approach (cf. Lem. 2.1 in Ref. [82]).

A physical advantage of introducing an invariant probability density $\rho$ in this manner is that it allows one to conceptually separate the vector field governing the evolution of the hypersurface—namely $V$—from the 'amount of probability' contained in each one of its 'regions' at different 'times'—the scalar field $\rho$. Indeed, we will see below that, while the Eulerian picture alone in principle only needs the vector field $J$ for its full formulation, the Lagrangian picture indirectly relies on Eq. (3.3) (via the map $\iota$).

The second question above has a more straightforward answer: The following proposition shows that the evolution does indeed respect the transversality condition.

**Proposition 3.1** *Let $\mathcal{Q}$ be a manifold and let $V$ be a vector field on $\mathcal{Q}$ that is nowhere tangent to an embedded submanifold $(\mathcal{S}_0, \iota_0)$ in $\mathcal{Q}$. Further, let $\Phi$ be the flow of $V$ and let $\tau$ be any real number such that the image $\mathcal{S}_\tau = \Phi_\tau\big(\iota_0(\mathcal{S}_0)\big)$ is defined.*

*Then, $\mathcal{S}_\tau$ is an embedded submanifold of $\mathcal{Q}$ that is nowhere tangent to $V$.* ◇

On the contrary, the following example shows that an initially spacelike hypersurface need not remain spacelike under the flow of a future-directed timelike vector field—even in the absence of curvature (cf. Remark 2.1 above).

*Example 3.1* In Minkowski 3-spacetime, standard coordinates $(ct, x, y)$ thereon, and for some $\omega > 0$ consider the observer (vector) field $V$ with values

$$\vec{V}(ct, x, y) = \omega \begin{pmatrix} 0 & -1 \\ 1 & 0 \end{pmatrix} \cdot \begin{pmatrix} x \\ y \end{pmatrix}, \tag{3.5a}$$

$$V^0(ct, x, y) = c\sqrt{1 + \Big(\vec{V}(t, x, y)/c\Big)^2}. \tag{3.5b}$$

Physically, this vector field describes a rotation about the observer curve $\tau \mapsto (c\tau, 0, 0)$ in $\mathbb{R}^3$. As Eq. (3.5a) is the equation of a linear vector field (cf. §1.1 in [86] and Ex. 3.2.8 in [177]) and the respective matrix factor is in the Lie algebra of the 2-dimensional rotation group, we find by exponentiation that

$$\Phi_\tau(t_0, x_0, y_0) = \begin{pmatrix} \sqrt{c^2 + \omega^2(x_0^2 + y_0^2)}\, \tau + ct_0 \\ x_0 \cos(\omega\tau) - y_0 \sin(\omega\tau) \\ x_0 \sin(\omega\tau) + y_0 \cos(\omega\tau) \end{pmatrix}. \tag{3.5c}$$

Consider now the initial hypersurface $\mathcal{S}_0 = \mathbb{R}^2$ with

$$\iota_0 \colon \mathcal{S}_0 \to \mathbb{R}^3 \colon (x_0, y_0) \mapsto \iota_0(x_0, y_0) = (0, x_0, y_0). \tag{3.5d}$$

Upon setting $r_0 := \sqrt{x_0^2 + y_0^2}$, the unit radial tangent vector at $(x_0, y_0) \neq (0, 0)$ that points away from the origin in $\mathcal{S}_0$ is given by

$$\hat{r}_{(x_0,y_0)} = \frac{\partial}{\partial r}\bigg|_{(x_0,y_0)} = \frac{1}{r_0}\begin{pmatrix} x_0 \\ y_0 \end{pmatrix}. \tag{3.5e}$$

58 An example of a choice of $\rho$ that is inappropriate in this sense is indeed quite common in the literature—so common that it has even found its way into undergraduate reference works such as Ref. [218] (cf. p. 141 therein): The choice $J^0 = \rho\, c$ of the 0th component of $J$ with respect to the standard coordinates in Minkowski spacetime. This does indeed allow for a one-to-one correspondence between the relativistic expression in Eq. (2.5a) and the respective non-relativistic one when integrated over any hypersurface of constant coordinate time $t$. But the coordinate-dependence of this choice is exposed, once one chooses other coordinates $\vec{x}'$ on the hypersurface obtained from a non-trivial Lorentz boost $\Lambda$ via $x' = \Lambda \cdot x$.

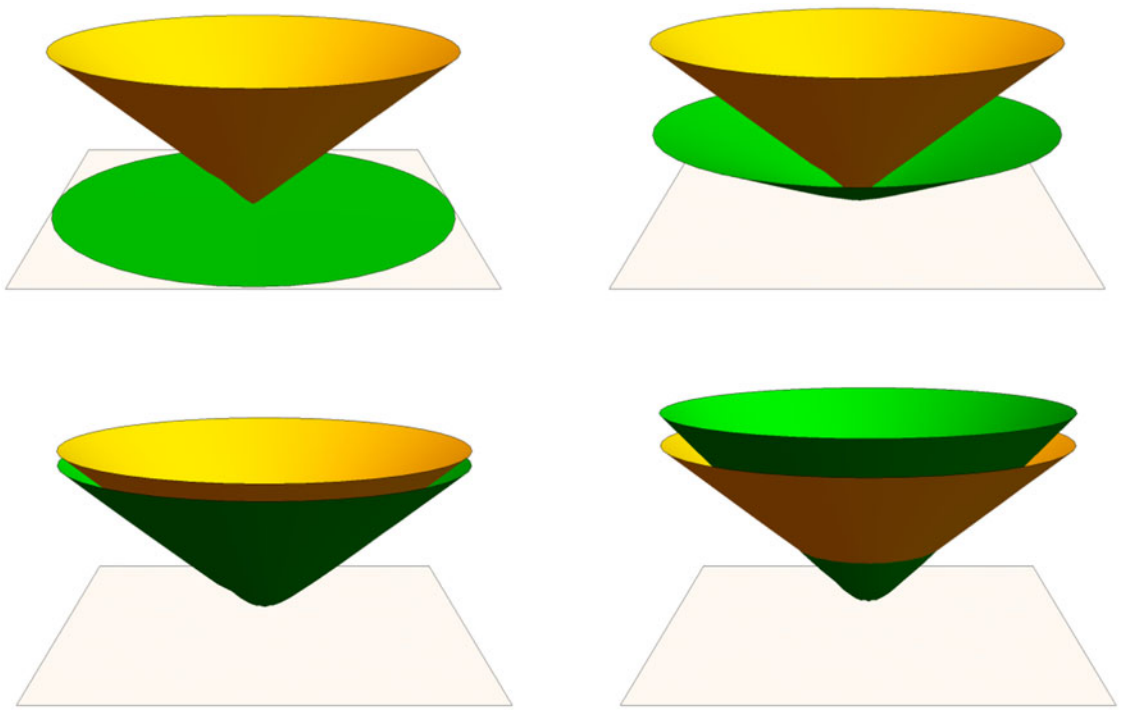

**Fig. 4** For Example 3.1 this figure illustrates the qualitative temporal evolution of a circular, green disk, initially lying in the $(t = 0)$-hypersurface $\iota_0(\mathcal{S}_0)$. Subsequent time steps are shown from left to right and from top to bottom. For each time step the white plane represents $\iota_0(\mathcal{S}_0)$. The yellow future light cone of the observer curve $\tau \mapsto (c\tau, 0, 0)$ moves upwards along the $ct$-axis, its tip always touching the disk. In the first three time steps, the disk bends towards the light cone, but does not intersect it—it stays spacelike. In the last time step, however, the disk has crossed the light cone—its causal character has changed

A calculation reveals that the pushforward of $\hat{r}_{(x_0,y_0)}$ under the map

$$\iota(\tau, \,.\,) = \Phi_\tau \circ \iota_0 \colon \mathcal{S}_0 \to \mathbb{R}^3 \colon (x_0, y_0) \mapsto \iota(\tau, x_0, y_0) = \Phi_\tau(0, x_0, y_0) \tag{3.5f}$$

with $\tau \in \mathbb{R}$ becomes lightlike at proper time

$$\tau = \frac{\sqrt{c^2 + \omega^2 r_0^2}}{\omega^2 r_0} \tag{3.5g}$$

and stays timelike afterwards. Therefore, the hypersurface $\mathcal{S}_\tau = \iota(\tau, \mathcal{S}_0)$ is spacelike for $\tau = 0$ and fails to be spacelike for any $\tau > 1/\omega$. The qualitative time evolution of $\mathcal{S}_0$ is illustrated in Fig. 4. ◇

The behavior exhibited in Example 3.1 seems to be generic. Indeed, we recently gave another example of a flow on a curved spacetime (Ex. 3 in Ref. [171]), that also does not preserve the spacelikeness of the initial hypersurface.

## 3.2 The Lagrangian picture

Having the central ideas of the Lagrangian picture in place, we proceed with the development of the rigorous theory. First, we define a suitable class of domains for the above map $\iota$.

**Definition 3.1** A *flow domain* $\mathcal{S}$ *for a manifold* $\mathcal{S}_0$ is an open subset of the product $\mathbb{R} \times \mathcal{S}_0$ such that for every $q$ in $\mathcal{S}_0$ the set

$$\mathcal{I}_q = \left\{ \tau \in \mathbb{R} \,\middle|\, (\tau, q) \in \mathcal{S} \right\} \tag{3.6}$$

is an open interval (cf. p. 211 in Ref. [127]). ◇

Of course, the domain of any flow of a vector field on $\mathcal{S}_0$ will be a flow domain.

Yet for our purposes we are interested in the case that $(\mathcal{S}_0, \iota_0)$ is an embedded submanifold of a manifold $\mathcal{Q}$ and that $\Phi$ is the (maximal) flow of a vector field $V$. In that instance, a flow domain is obtained by taking the (maximal) domain $\mathcal{S}$ on which the map

$$\iota \colon \mathcal{S} \to \mathcal{Q} \colon (\tau, q) \mapsto \iota(\tau, q) = \Phi_\tau\big(\iota_0(q)\big) \tag{3.7}$$

is well-defined.

The conceptual advantage of defining flow domains for $\mathcal{S}_0$ intrinsically, i.e. independent of any flow on $\mathcal{Q}$, is that, in the case that the above map $\iota$ is not surjective, one can forget about the values of $\Phi$ that do not lie in the image $\iota(\mathcal{S})$. Those points are not of relevance to the temporal evolution of the hypersurface. As noted in Sect. 1.3, this intrinsic approach turns out to be of fundamental importance for the many-body generalization.

Accordingly, in the next step, we provide a definition that captures the intrinsic properties of the tuple $(\mathcal{S}, \iota)$. It is at this point that the transversality condition appears in the Lagrangian picture.

**Definition 3.2** Let $\mathcal{Q}$ be a manifold and let $(\mathcal{S}_0, \iota_0)$ be an embedded submanifold of $\mathcal{Q}$.

(1) *A flowout from* $(\mathcal{S}_0, \iota_0)$ is a tuple $(\mathcal{S}, \iota)$ such that

(i) $\mathcal{S}$ is a flow domain for $\mathcal{S}_0$,

(ii) the map

$$\iota\colon \mathcal{S} \to \mathcal{Q}\colon (\tau, q) \mapsto \iota(\tau, q) \tag{3.8a}$$

satisfies $\iota_0 = \iota(0, \,.\,)$, and

(iii) $(\mathcal{S}, \iota)$ is an embedded submanifold of $\mathcal{Q}$.

(2) If $V$ is a vector field on $\mathcal{Q}$ that is nowhere tangent to $(\mathcal{S}_0, \iota_0)$, then the tuple $(\mathcal{S}, \iota)$ is called *a flowout from* $(\mathcal{S}_0, \iota_0)$ *along* $V$, if $(\mathcal{S}, \iota)$ is a flowout from $(\mathcal{S}_0, \iota_0)$ and we have

$$V_{\iota(\tau,q)} = \iota_* \left.\frac{\partial}{\partial \tau}\right|_{(\tau,q)} \tag{3.8b}$$

for all $(\tau, q) \in \mathcal{S}$.

The respective flowout is *maximal*, if there does not exist another flowout $\big(\mathcal{S}', \iota'\big)$ from $(\mathcal{S}_0, \iota_0)$ along $V$ such that $\mathcal{S} \subset \mathcal{S}'$.

◇

It is a simple consequence of the above definition that the vector field $\iota_*(\partial/\partial\tau)$ is nowhere tangent to $\big(\{\tau\} \times \mathcal{S}_0, \iota(\tau, \,.\,)\big)$ for all $\tau \in \mathbb{R}$ such that $\operatorname{dom} \iota(\tau, \,.\,) = \mathcal{S}_0$. As shown in the proof of Theorem 2.1.(1), this means that a hypersurface $(\mathcal{S}_0, \iota_0)$ in $\mathcal{Q}$ can only admit a flowout, if it is orientable.

*Remark 3.1* Three points are worthy of note with regards to Definition 3.2:

(1) In Ref. [127] Lee does not require flowouts $(\mathcal{S}, \iota)$ to be embedded.
His flowout theorem states the following (Thm. 9.20 in Ref. [127]): Given a vector field $V$ on a manifold $\mathcal{Q}$ that is nowhere tangent to an embedded submanifold $(\mathcal{S}_0, \iota_0)$ of $\mathcal{Q}$ and has flow $\Phi$, it is always possible to find a (not necessarily maximal) flow domain $\mathcal{S}$ for $\mathcal{S}_0$ such that the map $\iota$ in Eq. (3.7) is an injective immersion. That is, $(\mathcal{S}, \iota)$ is only an 'immersed' submanifold, not an embedded one.
While $\iota$ is indeed a smooth embedding whenever $(\mathcal{S}_0, \iota_0)$ is a hypersurface (point (d) in the above theorem), we conjecture that the general result may be strengthened in this manner regardless of dimension. To our knowledge, however, no such theorem exists in the literature. Despite this, Definition 3.2 does implicitly rely on this more general statement.

(2) There do indeed exist 'non-embedded flowouts'. An example of such a 'non-embedded flowout' from a 0-dimensional submanifold is given in Ex. 4.20 of Ref. [127]. In that instance the image $\iota(\mathcal{S})$ is dense in the ambient manifold $\mathcal{Q}$.
Nonetheless, it does not constitute a counterexample to the conjecture in point (1) above, because an 'embedded flowout' may be obtained via a restriction in domain $\mathcal{S}$ of the map $\iota$. The example does show, however, that a maximal 'embedded flowout' need not always exist.

(3) Definition 3.2 does not adequately account for the case that there exists a periodic integral curve of $V$ that intersects $(\mathcal{S}_0, \iota_0)$ (which would imply that the spacetime is 'vicious', cf. Sec. 4.2 in Ref. [146]):
If such a curve exists and $\Phi$ is the maximal flow of $V$, then the map $\iota$ in Eq. (3.7) will not be injective on its maximal domain $\mathcal{S}$. While the existence of such timelike/causal loops is often viewed as physically problematic (cf. Ref. [147] and p. 407 in Ref. [153]), Definition 3.2 is of purely mathematical nature and should therefore account for the possibility of periodic integral curves.
In Appendix C below we show how Definition 3.2 ought to be generalized to this case.
While it would be worthwhile to consider the theory in this article in this more general context, it would also make the mathematical relation between the Lagrangian and the Eulerian picture more subtle. In particular, Lemma 3.1 below would have to be turned into a local statement. We shall therefore abstain from using this more general definition here.

◇

It should be clear from the discussion in the beginning of this section that for the purposes of relativistic quantum theory we need to further constrain the classes of flowouts. Indeed, doing so constitutes the first major step in the construction of the Lagrangian picture.

**Definition 3.3** Let $(\mathcal{Q}, g, \mathcal{O})$ be a spacetime and let $(\mathcal{S}_0, \iota_0)$ be an orientable hypersurface in $\mathcal{Q}$.
A *future directed timelike 1-body flowout (from* $(\mathcal{S}_0, \iota_0)$*)* is a flowout $(\mathcal{S}, \iota)$ from $(\mathcal{S}_0, \iota_0)$ such that

$$\mathcal{V} = \iota_* \frac{\partial}{\partial \tau} \tag{3.9}$$

is a future-directed timelike vector field over $\mathcal{S}$.
Analogously, we define *future-directed lightlike* and *future-directed causal* 1*-body flowouts* by requiring $\mathcal{V}$ to have the respective property. ◇

While it is in principle possible to consider past-directed timelike/causal/lightlike 1-body flowouts in Definition 3.3 instead, because 'time flows forward' this would constitute an unconventional choice. Yet defining 'spacelike 1-body flowouts' would be physically inadequate, since a flowout for which $\mathcal{V}$ is spacelike cannot describe any temporal evolution of a 1-body quantum system.
The following proposition provides a general existence result for 1-body flowouts.

**Proposition 3.2** *In any spacetime, the following holds:*

(1) *There exists a future-directed timelike/causal/lightlike* 1*-body flowout.*
(2) *Given a future-directed timelike/causal/lightlike vector field* $V$ *on the spacetime and a hypersurface* $(\mathcal{S}_0, \iota_0)$ *such that* $V$ *is nowhere tangent, there exists a future-directed timelike/causal/lightlike* 1*-body flowout from* $(\mathcal{S}_0, \iota_0)$ *along* $V$.
(3) *If* $V_0$ *is a nowhere tangent, future-directed timelike vector field over a hypersurface* $(\mathcal{S}_0, \iota_0)$ *in the spacetime, then there exists a future-directed timelike vector field* $V$ *in a neighborhood of* $\iota_0(\mathcal{S}_0)$ *and a future-directed timelike* 1*-body flowout* $(\mathcal{S}, \iota)$ *from* $(\mathcal{S}_0, \iota_0)$ *along* $V$ *such that for all* $q$ *in* $\mathcal{S}_0$ *we have*

$$(V_0)_q = V_{\iota_0(q)} = V_{\iota(0,q)}. \tag{3.10}$$

◇

Point (3) of Proposition 3.2 essentially states that it is always possible to extend the construction in Sect. 2 to a kinematic description, at least in the timelike case.[59]

[59] We conjecture that this result also applies in the lightlike case. It might be possible to prove it by constructing a geodesic vector field $V$ that extends $V_0$ to a neighborhood of $\iota_0(\mathcal{S}_0)$.

The second major step in the construction of the Lagrangian picture is to show that one may think of the flow domain $\mathcal{S}$ as a spacetime in its own right.

**Lemma 3.1** *Let $(\mathcal{Q}, g, \mathcal{O})$ be a spacetime and let $(\mathcal{S}, \iota)$ be a future-directed timelike/causal/lightlike* 1*-body flowout. Further, let $\tilde{\iota}$ be the restriction of $\iota$ in codomain to its image in $\mathcal{Q}$.*

*Then $\big(\mathcal{S}, \iota^* g, (\tilde{\iota}^{-1})_* \mathcal{O}\big)$ is a spacetime isomorphic to $\big(\iota(\mathcal{S}), g\restriction_{\iota(\mathcal{S})}, \mathrm{Fr}(\iota(\mathcal{S})) \cap \mathcal{O}\big)$.*[60] ◇

Because of Lemma 3.1, one may 'forget' about the spacetime $\mathcal{Q}$ and 'carry' the physical description over to $\mathcal{S}$ instead, at least for the purpose of the kinematic Born rule. Indeed, one may freely switch between the two descriptions via the diffeomorphism $\tilde{\iota}$.

It is therefore justified to ask for the probability that the body is located in a 'region' $U$ in a certain hypersurface in $\mathcal{S}$, as opposed to a 'region' in an admissible hypersurface in $\mathcal{Q}$. This is indeed how we will formulate the kinematic 1-body Born rule in the Lagrangian picture.

*Remark 3.2* If $\iota(\mathcal{S})$ is properly contained in $\mathcal{Q}$, then the spacetime $\mathcal{S}$ in Lemma 3.1 may have different causal properties. It is, for instance, not possible for $\mathcal{S}$ to contain causal loops, since that would contradict injectivity of $\iota$. We refer the reader to Remark 3.1.(3) and Appendix C. References [146,147] provide a general discussion of the causal properties of spacetimes. ◇

While the discussion in the introduction of this section and Fig. 3 do explain the central ideas of the kinematic description, they are also potentially misleading: They implicitly suggest that the hypersurfaces $\{\tau\} \times \mathcal{S}_0 \subset \mathcal{S}$ with $\tau \in \mathbb{R}$ are the only ones of interest in the Lagrangian picture (or, equivalently, their images in $\mathcal{Q}$ under $\iota$). This would, however, indirectly contradict the general principle of relativity. Just as one should be able to freely choose the initial hypersurface $(\mathcal{S}_0, \iota_0)$ in $\mathcal{Q}$, hypersurfaces of the type $\{\tau\} \times \mathcal{S}_0$ ought not to hold any particular physical significance.

In order to mathematically account for this insight, we introduce the concept of a 'timeshift'.

**Definition 3.4** Let $(\mathcal{S}, \iota)$ be a flowout from a submanifold $(\mathcal{S}_0, \iota_0)$ in a manifold $\mathcal{Q}$.

We call a (smooth) function $\mathcal{T}$ on $\mathcal{S}_0$ such that its graph $\mathrm{graph}(\mathcal{T})$ is contained in $\mathcal{S}$ a *(global) shift (of $(\mathcal{S}_0, \iota_0)$ in $\mathcal{S}$)*. Furthermore, if $U$ is any open subset of $\mathcal{S}_0$, we call a map $\mathcal{T} : U \to \mathbb{R}$ with $\mathrm{graph}(\mathcal{T}) \subset \mathcal{S}$ a *local shift (of $(\mathcal{S}_0, \iota_0)$ in $\mathcal{S}$)*.

In case $(\mathcal{S}, \iota)$ is a future-directed timelike/causal/lightlike 1-body flowout, then the shift $\mathcal{T}$ is called a *(global* 1*-body) timeshift (of $(\mathcal{S}_0, \iota_0)$ in $\mathcal{S}$)*. Analogously, in this context a local shift $\mathcal{T}$ is called a *local (*1*-body) timeshift*. ◇

The general idea of a shift $\mathcal{T}$ is that it induces a mapping

$$\iota_{\mathcal{T}} : \mathcal{S}_0 \to \mathcal{S} : q \mapsto \iota_{\mathcal{T}}(q) = \iota\big(\mathcal{T}(q), q\big) \tag{3.11}$$

which 'shifts' the submanifold $(\mathcal{S}_0, \iota_0)$ to the submanifold $(\mathcal{S}_0, \iota_{\mathcal{T}})$ in $\mathcal{Q}$. The use of local shifts is convenient, if one does not want to 'shift the whole submanifold'.

Accordingly, in the Lagrangian picture (global) timeshifts model the time evolution of the relativistic 1-body system: For $\tau \in \mathbb{R}$ the hypersurface $\{\tau\} \times \mathcal{S}_0$ in $\mathcal{S}$ is obtained by choosing the particular timeshift $\mathcal{T}$ with values $\mathcal{T}(q) = \tau$ for all $q \in \mathcal{S}_0$—but in the general case, there is no requirement for $\mathcal{T}$ to be constant.[61] An illustration is given in Fig. 5.

Having gathered all the main ingredients of the Lagrangian picture, we may now state the respective central theorem.

[60] For a manifold $\mathcal{M}$ we denote its frame bundle by $\mathrm{Fr}(\mathcal{M})$. If $n = \dim \mathcal{M}$, then $\mathrm{Fr}(\mathcal{M})$ is a principal $\mathrm{GL}(\mathbb{R}^n)$-bundle over $\mathcal{M}$.

[61] By excluding closed sets of measure zero on $\mathcal{S}_0$, it is even possible to consider (local) timeshifts that would otherwise be discontinuous.

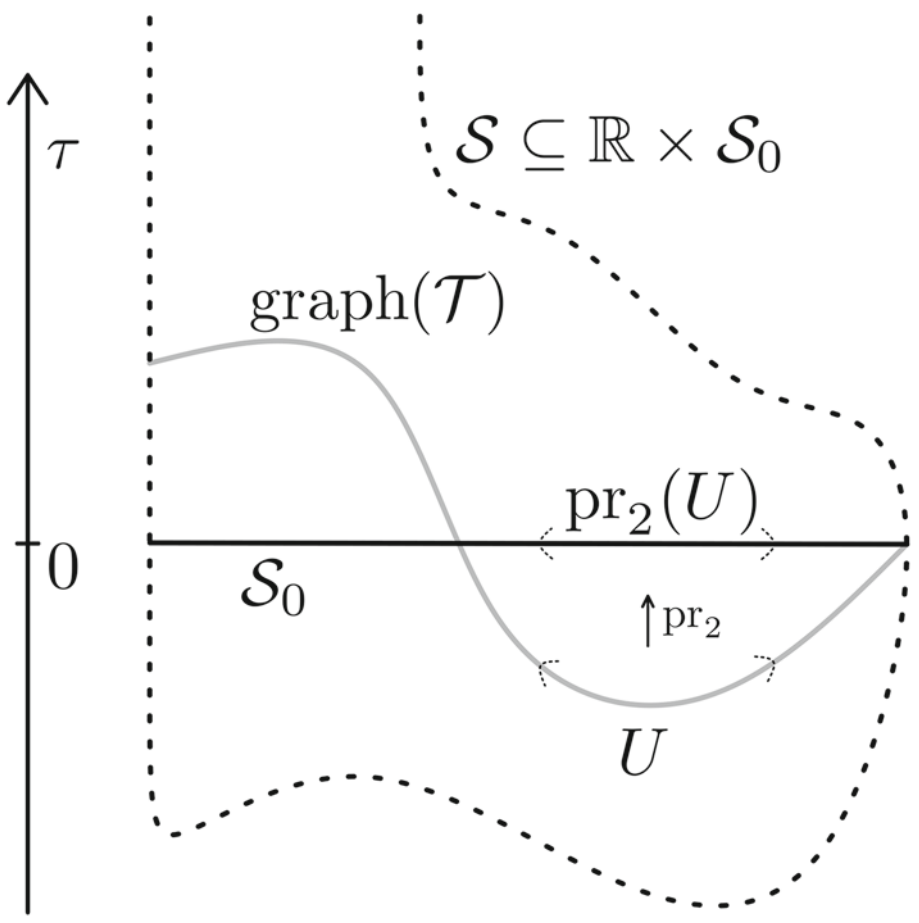

**Fig. 5** This sketch shows how the Born rule is formulated in the Lagrangian picture.(Note that we simplified the notation by identifying the set $\{0\} \times \mathcal{S}_0$ with $\mathcal{S}_0$ and the map $\big(0, \mathrm{pr}_2(\,.\,)\big)$ with $\mathrm{pr}_2$.) The initial hypersurface is $\mathcal{S}_0$, given as the subset $\{0\} \times \mathcal{S}_0$ in the depiction. The respective flow domain $\mathcal{S} \subseteq \mathbb{R} \times \mathcal{S}_0$ is enclosed by the dashed line and, in this instance, partially unbounded in positive $\tau$-direction. Following Lemma 3.1, the general idea is that the manifold $\mathcal{S}$ allows one to reformulate the dynamics in $\mathcal{Q}$ such that said dynamics becomes trivial (it is implicitly encoded in the map $\iota\colon \mathcal{S} \to \mathcal{Q}$). As in the instantaneous case, we formulate the Born rule on (the image of) an embedded hypersurface—yet in the Lagrangian picture, this hypersurface is the graph of a (global) timeshift $\mathcal{T}$, shown here in gray. If one has probability conservation, then for a given 'region' $U \subseteq \mathrm{graph}(\mathcal{T})$ the probability that the body is located in a subset $\iota(U)$ of $\iota(\mathrm{graph}(\mathcal{T}))$ is simply the probability to that the body is located in the subset $\iota_0\left(\mathrm{pr}_2(U)\right)$ of $\iota_0\left(\mathcal{S}_0\right)$ (cf. Theorem 3.1.(5))

**Theorem 3.1** *Let $(\mathcal{Q}, g, \mathcal{O})$ be a spacetime. Let $(\mathcal{S}, \iota)$ be a future-directed timelike/causal/lightlike* 1*-body flowout from a hypersurface $(\mathcal{S}_0, \iota_0)$ in $\mathcal{Q}$.*

*Furthermore, let $\mathcal{T}$ be a (global) timeshift of $\mathcal{S}_0$ in $\mathcal{S}$ and define the function*

$$\mathcal{T}'\colon \mathcal{S}_0 \to \mathcal{S}\colon q \mapsto \mathcal{T}'(q) = (\mathcal{T}(q), q). \tag{3.12a}$$

*The following holds:*

(1) (i) $\left(\mathcal{S}_0, \mathcal{T}'\right)$ *is an embedded submanifold of $\mathcal{S}$ that is equivalent to* $\mathrm{graph}(\mathcal{T})$ *(together with its natural inclusion in $\mathcal{S}$), and*

(ii) $\mathrm{graph}(\mathcal{T})$ *is orientable and carries a canonical orientation.*

(2) *Denote by $\mu$ the canonical volume form on $\mathcal{Q}$. Then*

$$\upsilon = \frac{1}{c}\,\frac{\partial}{\partial \tau} \lrcorner\, \iota^* \mu \tag{3.12b}$$

*restricts to a volume form on* $\mathrm{graph}(\mathcal{T})$*. Moreover, $\left(\mathcal{T}'\right)^* \upsilon$ is a volume form on $\mathcal{S}_0$.*

(3) *Choose any (smooth) strictly positive scalar field $\varrho$ on $\mathcal{S}$. Then there exists a unique scalar field $\mathring{\varrho}$ on $\mathcal{S}$ such that*[62]

$$\frac{\partial}{\partial \tau}\,(\varrho\, \upsilon) = \mathring{\varrho}\, \upsilon. \tag{3.12c}$$

---

[62] By definition, if $(\alpha_s)_{s \in \mathbb{R}}$ is a 1-parameter family of $k$-forms on a manifold $\mathcal{S}_0$, then for all $(s, q) \in \mathbb{R} \times \mathcal{S}_0$ and $Y_1, \ldots, Y_k \in \mathrm{T}_q \mathcal{S}_0$ one may define

$$\left(\frac{\partial \alpha_s}{\partial s}\right)_q (Y_1, \ldots, Y_k) = \frac{\partial}{\partial s}\left((\alpha_s)_q\,(Y_1, \ldots, Y_k)\right).$$

This definition extends to the case that $\alpha$ is only defined on an open subset $\mathcal{S}$ of $\mathbb{R} \times \mathcal{S}_0$.

(4) *Set* $\mathrm{graph}(0) = \{0\} \times \mathcal{S}_0 \subset \mathcal{S}$. *Let* $\varrho$ *as above satisfy*

$$\int_{\mathrm{graph}(0)} \varrho\, \upsilon = 1 \tag{3.12d}$$

*and* $\mathring{\varrho} = 0$. *Then*

$$\mathbb{P}_L \colon \mathcal{B}^*\big(\mathrm{graph}(\mathcal{T})\big) \to [0,1] \colon U \mapsto \mathbb{P}_L(U) = \int_U \varrho\, \upsilon \tag{3.12e}$$

*defines a probability measure on the measurable space* $\big(\mathrm{graph}(\mathcal{T}),\ \mathcal{B}^*(\mathrm{graph}(\mathcal{T}))\big)$. *Moreover, for* $U \in \mathcal{B}^*(\mathrm{graph}(\mathcal{T}))$ *the probability* $\mathbb{P}_L(U)$ *is* 0 *if and only if* $U$ *is a Lebesgue null set.*

(5) *Consider the projection mapping*

$$\mathrm{pr}_2 \colon \mathcal{S} \to \mathcal{S}_0 \colon (\tau, q) \mapsto \mathrm{pr}_2(\tau, q) = q\,. \tag{3.12f}$$

*If* $\varrho$ *satisfies the properties of point (4) above, then for all* $U \in \mathcal{B}^*(\mathrm{graph}(\mathcal{T}))$ *we have*

$$\mathbb{P}_L(U) = \int_{\mathrm{pr}_2(U)} \varrho(0,\,.\,)\, \upsilon_{(0,\,.\,)} = \mathbb{P}_L\big(0'(\mathrm{pr}_2(U))\big). \tag{3.12g}$$

*In particular,* $\mathbb{P}_L(U)$ *only depends on* $\mathrm{pr}_2(U)$, *which is independent of the choice of* $\mathcal{T}$. ◇

Point (1) of Theorem 3.1 is a consistency result. Point (2) shows that $\upsilon$ may be viewed as a ‘spatial volume form’. Point (3) motivates the introduction of the *Langrangian continuity equation*, that is the continuity equation in the Lagrangian picture:

$$\mathring{\varrho} = 0. \tag{3.13}$$

Moreover, Eq. (3.12c) identifies $\mathring{\varrho}$ as a geometric invariant—which is of particular relevance whenever one lifts the requirement of probability conservation (cf. Sect. 4). Point (4) states that the kinematic 1-body Born rule in the Langrangian picture, Eq. (3.12e), does indeed yield a probability in the strict mathematical sense—that is probability conservation holds—whenever the Lagrangian continuity equation is satisfied. Regarding the relation of point (4) to the general principle of relativity we refer to Sect. 1.2 and Footnote 28 above. If probability conservation holds, then point (5) provides us with a simple way to compute the respective probability.

*Remark 3.3* The reader may have noticed that the notation $\mathbb{P}_\mathrm{L}(U)$ for the respective probability does not make any explicit reference to a choice of timeshift. There is indeed a mathematical justification for this notation:

Consider the set

$$\mathcal{B}^*_\mathrm{L}(\mathcal{S}) = \big\{U \subset \mathcal{S} \big| \exists \text{ local timeshift } \mathcal{T} \colon\ U \in \mathcal{B}^*\big(\mathrm{graph}(\mathcal{T})\big)\big\}\,. \tag{3.14}$$

Since for $U \in \mathcal{B}^*\big(\mathrm{graph}(\mathcal{T})\big)$ the integral in Eq. (3.12e) does not depend on the values of the timeshift $\mathcal{T}$ on $\mathrm{graph}(\mathcal{T}) \setminus U$, we may understand $\mathbb{P}_\mathrm{L}$ as a map from $\mathcal{B}^*_\mathrm{L}(\mathcal{S})$ to the interval [0, 1]. The downside of this approach is that $\mathbb{P}_\mathrm{L}$ then ceases to be a probability measure in the mathematical sense, because $\mathcal{B}^*_\mathrm{L}(\mathcal{S})$ does not contain $\mathcal{S}$ itself and is thus not a $\sigma$-algebra.

Indeed, one may pursue this approach further and even define respective ‘Lagrangian $L^p$-spaces’ (cf. Remark 2.2.(4)). Since we only treat the smooth case here, however, we shall abstain from a more detailed treatment thereof.

Also note that the Lagrangian continuity equation (3.13) does not need to hold for the above definition of $\mathbb{P}_\mathrm{L}$ to be sensible. ◇

The next lemma provides local coordinate expressions for $\mathring{\varrho}$ and $\mathbb{P}_{\mathrm{L}}$.

**Lemma 3.2** *Consider the situation of Theorem 3.1 above, and let $(V, \xi)$ be an oriented chart on $\mathcal{S}_0$.*

(1) *With respect to the coordinates $(\tau, \xi)$ on $\mathcal{S}$ we have*[63]

$$\mathring{\varrho}(\tau, \xi) = \frac{1}{\sqrt{-\det(\iota^* g)}(\tau, \xi)} \frac{\partial}{\partial \tau} \left( \varrho(\tau, \xi) \sqrt{-\det(\iota^* g)}(\tau, \xi) \right). \tag{3.15a}$$

(2) *Let $\mathcal{T}$ be a local timeshift of $\mathcal{S}_0$ in $\mathcal{S}$ with $\operatorname{dom} \mathcal{T} \cap V \neq \varnothing$. Then for all $W \in \mathcal{B}^*(V)$ we have*

$$\mathbb{P}_L\big(\mathcal{T}'(W)\big) = \frac{1}{c} \int_W \varrho(\mathcal{T}(\xi), \xi) \sqrt{-\det(\iota^* g)}(\mathcal{T}(\xi), \xi)\, \mathrm{d}^n \xi. \tag{3.15b}$$

*Moreover, if $\mathring{\varrho}(\tau, \xi) \equiv 0$, we may set $\mathcal{T}(\xi) \equiv 0$ in the above expression without changing the result (cf. Theorem 3.1.(5)).*

◇

Equation (3.15a) shows that the Lagrangian continuity equation (3.13) is easily solved locally.

In fact, locally, the Lagrangian picture may be viewed as a choice of local slice coordinates for $(\mathcal{S}_0, \iota_0)$ in $\mathcal{Q}$ in which the vector field $V$ is simply $\partial/\partial\tau$. This is, however, a simplified view, for the construction provided here is indeed a global one. Moreover, this view fails to acknowledge that the Lagrangian picture opens up a potentially viable path to the many-body generalization.

In order to give an example of how the Lagrangian picture is employed in practice, we continue with Example 3.1 above.

*Example 3.2* By construction, the tuple $(\mathcal{S}, \iota)$ with $\mathcal{S} = \mathbb{R}^3$ and $\iota$ determined by Eq. (3.5f) is a maximal future-directed timelike 1-body flowout from the hypersurface $(\mathcal{S}_0, \iota_0)$ along the vector field $V$. In particular, we find

$$\mathcal{V}_{(\tau, x_0, y_0)} \equiv V_{\iota(\tau, x_0, y_0)} = \begin{pmatrix} c\sqrt{1 + (\omega r_0/c)^2} \\ \omega \begin{pmatrix} -\sin(\omega\tau) & -\cos(\omega\tau) \\ \cos(\omega\tau) & -\sin(\omega\tau) \end{pmatrix} \cdot \begin{pmatrix} x_0 \\ y_0 \end{pmatrix} \end{pmatrix}. \tag{3.16a}$$

A calculation reveals that the values of $\upsilon$ are given by

$$\upsilon_{(\tau, x_0, y_0)} = \sqrt{1 + (\omega r_0/c)^2}\, \mathrm{d}x_0 \wedge \mathrm{d}y_0. \tag{3.16b}$$

Since a function $\varrho$ satisfies the Lagrangian continuity equation whenever $\varrho\, \upsilon$ is independent of the parameter $\tau$, one has probability conservation whenever $\varrho$ itself is independent of $\tau$. Accordingly, we may choose a constant

[63] If we denote the coordinate representation of $\iota$ by $(\tau, \xi) \mapsto \kappa(\tau, \xi)$ and its Jacobian determinant by

$$\det\left(\frac{\partial \kappa}{\partial(\tau, \xi)}\right) = (n+1)!\, \frac{\partial \kappa^{[0}}{\partial \tau} \frac{\partial \kappa^1}{\partial \xi^1} \cdots \frac{\partial \kappa^{n]}}{\partial \xi^n},$$

then, by considering different local expressions for $\iota^*\mu$, one shows that for all $(\tau, \xi)$ one has

$$\sqrt{-\det(\iota^* g)}(\tau, \xi) = \det\left(\frac{\partial \kappa}{\partial(\tau, \xi)}\right)(\tau, \xi)\, \sqrt{-\det g}\big(\kappa(\tau, \xi)\big).$$

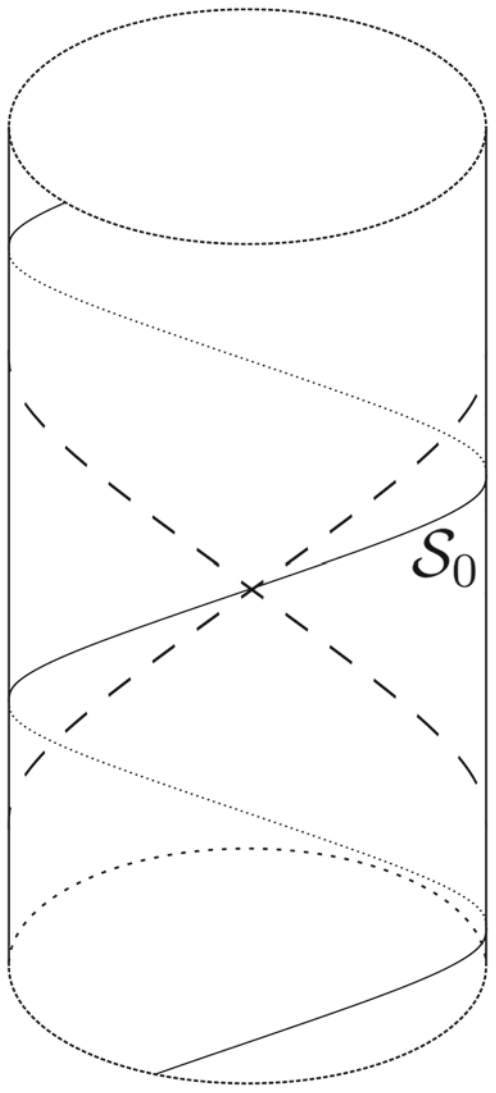

**Fig. 6** This figure depicts a cylinder spacetime and an unphysical choice of a spacelike hypersurface $\mathcal{S}_0$ therein. See Remarks 2.2.(2) and 3.4 for details

$\sigma > 0$ and a function $\varrho$ with values

$$\varrho(\tau, x_0, y_0) = \frac{1}{2\pi\sigma^2\sqrt{1+(\omega r_0/c)^2}}\, e^{-r_0^2/2\sigma^2} \tag{3.16c}$$

to define a Lagrangian probability measure $\mathbb{P}_{\mathrm{L}}$ on the graph of any global timeshift $\mathcal{T}\colon (x_0, y_0) \mapsto \mathcal{T}(x_0, y_0)$ of $\mathcal{S}_0$. ◇

*Remark 3.4* In order to obtain a one-to-one correspondence between the Eulerian picture and Lagrangian picture, in most physical situations one wants the map $\iota$ to be surjective (so that it becomes a diffeomorphism).

There are, however, numerous reasons why $\iota$ may fail to be surjective. We shall only mention some of them here:

There may be points on $\mathcal{Q}$ where the current density vector field vanishes. If one has probability conservation, one may safely exclude them by restricting $\iota$ to its image, as done in Lemma 3.1 above.

There may be topological reasons: Some of the points of the spacetime may be 'missing', for instance due to singularities of the metric. This might point towards a limitation of the given physical model. If the spacetime admits causal loops, there might not be any maximal flowout from any suitable hypersurface in $\mathcal{Q}$. In that case we refer to Remark 3.1.(3) and Appendix C.

Yet, $\iota$ may also fail to be surjective because of an inappropriate choice of initial hypersurface: Fig. 6 depicts such a hypersurface $\mathcal{S}_0$, given as a subset of a flat cylinder spacetime $\mathcal{Q} = \mathbb{R} \times \mathbb{S}^1$. $\mathcal{S}_0$ is chosen to be an inextendible spacelike geodesic. The long-dashed lines indicate the light cone at $(0, 0)$. Even though $\mathcal{S}_0$ is spacelike, connected, and even maximal, any future-directed causal 1-body flowout $(\mathcal{S}, \iota)$ from $\mathcal{S}_0$ will fail to be surjective—due to the injectivity condition imposed on $\iota$. There do, however, exist choices of initial (spacelike) hypersurfaces—such as $\mathcal{S}_0 = \{0\} \times \mathbb{S}^1$—for which any future-directed causal vector field $V$ on $\mathcal{Q}$ gives rise to a respective surjective 1-body flowout along $V$.

Even though $\mathcal{Q}$ is globally hyperbolic, $\mathcal{S}_0$ is, of course, not a Cauchy surface. The approach by Lienert and Tumulka [132] and Miller et al. [144] would therefore exclude this choice as well.

Irrespective of whether one takes Definition 3.2 or Definition C.1 as the underlying definition of a flowout, we propose that in general those initial hypersurfaces are admissible for which the map $\iota$ is a bijection—if such a choice exists. Ultimately, however, this condition can and ought not to replace good physical judgement. ◇

### 3.3 The Eulerian picture

As discussed in the beginning of this section, in the Eulerian picture the evolution of the initial hypersurface is indirectly determined by a future-directed timelike, causal, or lightlike vector field $V$ on the spacetime $\mathcal{Q}$. The main purpose of this subsection is to spell out analogue theorems to those stated in Sect. 3.2 and to discuss the precise relationship between the two pictures that goes beyond Lemma 3.1 above. Lemma 3.3 below shows how the kinematic Born rule in the Eulerian picture leads to the instantaneous Born rule of Sect. 2.

Accordingly, there are two main theorems in this subsection: Corollary 3.1 is the Eulerian analogue to Theorem 3.1 above. Theorem 3.2 relates the two pictures in full detail.

We shall first state Corollary 3.1. It is a consequence of Theorem 3.1 above and Theorem 3.2 below.

**Corollary 3.1** *Let $(\mathcal{Q}, g, \mathcal{O})$ be a spacetime, let $V$ be a future-directed timelike/causal/lightlike vector field on $\mathcal{Q}$ with flow $\Phi$, and let $(\mathcal{S}_0, \iota_0)$ be a hypersurface in $\mathcal{Q}$ such that $V$ is nowhere tangent and*

$$\mathcal{Q} = \left\{ \Phi_\tau\big(\iota_0(q)\big) \,\big|\, q \in \mathcal{S}_0 \text{ and } \big(\tau, \iota_0(q)\big) \in \operatorname{dom}\Phi \right\}. \tag{3.17a}$$

*Furthermore, let $(\mathcal{N}, \varphi)$ be a hypersurface such that for all $q$ in $\mathcal{S}_0$ the (maximal) integral curve $\tau \mapsto \Phi_\tau\big(\iota_0(q)\big)$ intersects $(\mathcal{N}, \varphi)$ transversally in exactly one point.*

*The following holds:*

(1) (i) *The image $\varphi(\mathcal{N})$ together with the natural inclusion is equivalent to $(\mathcal{N}, \varphi)$ as a hypersurface, and*
(ii) *$\varphi(\mathcal{N})$ is orientable and carries a canonical orientation.*

(2) *Denote by $\mu$ the canonical volume form on $\mathcal{Q}$. Then*

$$\nu = \frac{1}{c} V \lrcorner \mu \tag{3.17b}$$

*restricts to a volume form on $\varphi(\mathcal{N})$. Moreover, $\varphi^*\nu$ is a volume form on $\mathcal{N}$.*

(3) *Choose any (smooth) strictly positive scalar field $\rho$ on $\mathcal{Q}$ and denote by* div *the divergence with respect to $\mu$.*[64] *Then the following holds:*

$$\mathcal{L}_V\,(\rho\,\nu) = \operatorname{div}(\rho\, V)\ \nu. \tag{3.17c}$$

*In particular, if* $\operatorname{div}(\rho\, V)$ *vanishes everywhere, the form $\rho\,\nu$ is absolutely invariant with respect to $V$ (cf. Def. 4.2.15.3 in Ref.* [177]*).*

(4) *Let $\rho$, as above, satisfy*

$$\int_{\mathcal{S}_0} \rho\,\nu = 1 \tag{3.17d}$$

*and* $\operatorname{div}(\rho\, V) = 0$*. Then*

$$\mathbb{P}_E \colon \mathcal{B}^*\big(\varphi(\mathcal{N})\big) \to [0,1] \colon U \mapsto \mathbb{P}_E(U) = \int_U \rho\,\nu \tag{3.17e}$$

[64] By definition, for any vector field $Y$ on $\mathcal{Q}$ we have

$$\mathcal{L}_Y\,\mu = \operatorname{div}(Y)\,\mu.$$

*defines a probability measure on the measurable space* $\big(\varphi(\mathcal{N}), \mathcal{B}^*(\varphi(\mathcal{N}))\big)$*. Moreover, for* $U \in \mathcal{B}^*\big(\varphi(\mathcal{N})\big)$ *the probability* $\mathbb{P}_E(U)$ *is* 0 *if and only if* $U$ *is a Lebesgue null set.*

(5) *Let* $\mathcal{T} : \mathcal{S}_0 \to \mathbb{R}$ *be the (unique) map such that for all* $q \in \mathcal{S}_0$ *we have*

$$\Phi_{\mathcal{T}(q)}\big(\iota_0(q)\big) \in \varphi(\mathcal{N}). \tag{3.17f}$$

*For any* $q$ *in* $\mathcal{S}_0$ *set*

$$\iota_{\mathcal{T}}(q) = \Phi_{\mathcal{T}(q)}\left(\iota_0(q)\right). \tag{3.17g}$$

*If* $\rho$ *satisfies the properties of point (4) above, then for all* $U \in \mathcal{B}^*(\mathcal{S}_0)$ *we have*

$$\mathbb{P}_E\big(\iota_{\mathcal{T}}(U)\big) = \int_{\iota_0(U)} \rho\, \nu = \mathbb{P}_E\big(\iota_0(U)\big). \tag{3.17h}$$

*In particular,* $\mathbb{P}_E(\iota_{\mathcal{T}}(U))$ *only depends on* $U$ *and therefore not on the choice of hypersurface* $(\mathcal{N}, \varphi)$.

◇

Of course, Eq. (3.17e) defines the kinematic 1-body Born rule in the Eulerian picture and

$$\operatorname{div}(\rho V) = 0 \tag{3.18}$$

is the *Eulerian continuity equation*.

The main idea in showing the equivalence of the two pictures is that $\iota$ is a diffeomorphism whenever it is surjective. Moreover, even if it is not surjective, it may be made surjective by restricting it to its image, as done in Lemma 3.1. By applying the diffeomorphism $\iota$ accordingly, every statement in one picture may then be translated to the other. The commutative diagram in Fig. 7 shows how this is done for the evolution of the initial hypersurface.

**Theorem 3.2** *Let* $(\mathcal{Q}, g, \mathcal{O})$ *be a spacetime and let* $(\mathcal{S}_0, \iota_0)$ *be a hypersurface in* $\mathcal{Q}$.

(1) (i) *If* $V$ *is a future-directed timelike/causal/lightlike vector field on* $\mathcal{Q}$ *that is nowhere tangent to* $(\mathcal{S}_0, \iota_0)$*, then there exists a future-directed timelike/causal/lightlike* 1*-body flowout* $(\mathcal{S}, \iota)$ *from* $(\mathcal{S}_0, \iota_0)$ *along* $V$*. In particular, if* $\Phi$ *denotes the flow of* $V$*, then for all* $(\tau, q) \in \mathcal{S}$ *we have*

$$\iota(\tau, q) = \Phi_\tau\big(\iota_0(q)\big) \tag{3.19a}$$

*and*

$$\mathcal{V}_{(\tau,q)} = \iota_* \left.\frac{\partial}{\partial \tau}\right|_{(\tau,q)} = V_{\iota(\tau,q)}. \tag{3.19b}$$

(ii) *Conversely, if* $(\mathcal{S}, \iota)$ *is a future-directed timelike/causal/lightlike* 1*-body flowout from* $(\mathcal{S}_0, \iota_0)$*, then Eq. (3.19b) defines a future-directed timelike/causal/lightlike vector field* $V$ *on* $\mathcal{Q}$ *with flow* $\Phi$ *that is nowhere tangent to* $(\mathcal{S}_0, \iota_0)$ *and Eq. (3.19a) holds.*
*Furthermore, if* $\iota(\mathcal{S}) = \mathcal{Q}$*, then Eq. (3.17a) holds as well.*

*In the following, let either set of assumptions of point (1) be satisfied and assume in addition that* $\mathcal{Q} = \iota(\mathcal{S})$ *(so that* $\iota$ *becomes a diffeomorphism):*

(2) *If we define* $\nu$ *via Eq. (3.17b) and* $\upsilon$ *via Eq. (3.12b), then*

$$\iota^* \nu = \upsilon. \tag{3.19c}$$

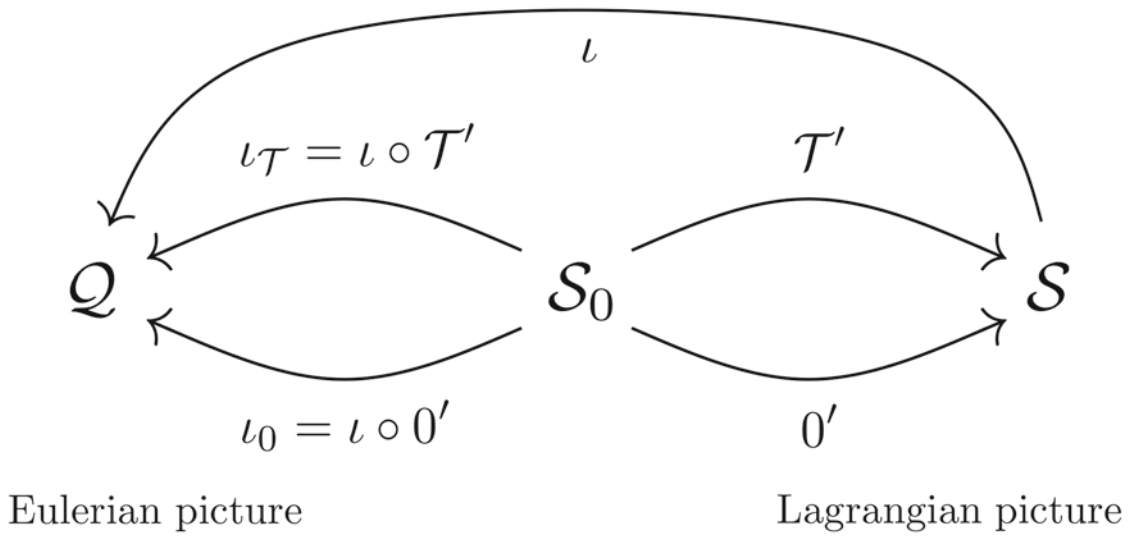

**Fig. 7** This commutative diagram provides an overview of how the time evolution of the initial hypersurface $(\mathcal{S}_0, \iota_0)$—or $(\mathcal{S}_0, 0')$, respectively—is modeled in either picture. In both pictures, a global timeshift $\mathcal{T}$ is associated with the 'evolved' hypersurface (cf. Definition 3.4 and Theorem 3.2.(4)): In the Lagrangian picture, this is the hypersurface $(\mathcal{S}_0, \mathcal{T}')$, with $\mathcal{T}'$ given by Eq. (3.12a), in the Eulerian picture, it is the hypersurface $(\mathcal{S}_0, \iota_{\mathcal{T}})$

(3) *Given either $\rho$ from Corollary 3.1.(3) or $\varrho$ from Theorem 3.1.(3), we define the other scalar field via the relation*

$$\rho \circ \iota = \varrho. \tag{3.19d}$$

*Then, we have the identity*

$$\iota^*\big(\mathrm{div}(\rho\, V)\, \nu\big) = \mathring{\varrho}\, \upsilon. \tag{3.19e}$$

*In particular, the following equivalence holds:*

$$\mathrm{div}\,(\rho\, V) = 0 \quad \Longleftrightarrow \quad \mathring{\varrho} = 0. \tag{3.19f}$$

(4) *Given a global timeshift $\mathcal{T}$ of $(\mathcal{S}_0, \iota_0)$ in $\mathcal{S}$, define $\iota_{\mathcal{T}}$ by Eq. (3.11) and set $\mathcal{S}_{\mathcal{T}} = \iota_{\mathcal{T}}(\mathcal{S}_0)$. Then for a hypersurface $(\mathcal{N}, \varphi)$ in $\mathcal{Q}$ the following are equivalent:*

(i) *For all $q$ in $\mathcal{S}_0$ the (maximal) integral curve $\tau \mapsto \Phi_\tau\big(\iota_0(q)\big)$ intersects $(\mathcal{N}, \varphi)$ transversally in exactly one point.*

(ii) *There exists a global timeshift $\mathcal{T}$ of $(\mathcal{S}_0, \iota_0)$ such that*

$$\varphi(\mathcal{N}) = \mathcal{S}_{\mathcal{T}}. \tag{3.19g}$$

(5) *If $\mathcal{T}$ is any local timeshift of $(\mathcal{S}_0, \iota_0)$ in $\mathcal{S}$, then for all $U \in \mathcal{B}^*(\mathcal{S}_0)$ we have*

$$\int_{\iota_{\mathcal{T}}(U)} \rho\, \nu = \int_{\mathcal{T}'(U)} \varrho\, \upsilon, \tag{3.19h}$$

*provided that either one of the respective integrals exists. In particular, if either one of the conditions of Eq. (3.19f) above is satisfied, we may write*

$$\mathbb{P}_E\big(\iota_{\mathcal{T}}(U)\big) = \mathbb{P}_L\big(\mathcal{T}'(U)\big) \tag{3.19i}$$

*(cf. Theorem 3.1.(4) and Corollary 3.1.(4)).* ◇

*Remark 3.5* In Remark 3.3, we argued that the map $\mathbb{P}_{\mathrm{L}}$ may be defined without referring to a particular choice of timeshift. The map $\mathbb{P}_{\mathrm{E}}$ may indeed be extended in an analogous manner:

Recall that for $q \in \mathcal{Q}$ one defines the *streamline* of the vector field $V$ at $q$ to be the image of the respective maximal integral curve starting at $q$. Since it is possible for two streamlines at different points of $\mathcal{Q}$ to coincide, one may view a streamline as an equivalence class of maximal integral curves of $V$. In this manner, we may refer to a streamline without needing to specify a particular point on $\mathcal{Q}$ or a particular integral curve. Furthermore, every streamline is an 'immersed' submanifold of $\mathcal{Q}$ (together with the natural inclusion and if equipped with the topology and smooth structure obtained from a respective integral curve).

Under the assumptions in the first sentence of Corollary 3.1, we may then define the set

$$\begin{aligned}\mathcal{B}^*_{\mathrm{E}}(\mathcal{Q}) = \Big\{U \subset \mathcal{Q} \,\Big|\, &\exists \text{ hypersurface } (\mathcal{N}, \varphi) \text{ in } \mathcal{Q}\colon \text{ every streamline of } V \\ &\text{intersects } (\mathcal{N}, \varphi) \text{ transversally in at most one point} \\ &\text{and } U \in \mathcal{B}^*\big(\varphi(\mathcal{N})\big)\Big\}. \end{aligned} \tag{3.20a}$$

A major benefit of this definition is that it does not make any explicit reference to $(\mathcal{S}_0, \iota_0)$ or $\iota_0(\mathcal{S}_0)$.

In full analogy to the construction in the Lagrangian picture, we may then extend the domain of $\mathbb{P}_{\mathrm{E}}$ to $\mathcal{B}^*_{\mathrm{E}}(\mathcal{Q})$.

Under the additional assumption that there exists a 1-body flowout $(\mathcal{S}, \iota)$ from $(\mathcal{S}_0, \iota_0)$ along $V$ such that $\iota$ is surjective, we may also use Theorem 3.2.(4) to show that

$$\mathcal{B}^*_{\mathrm{E}}(\mathcal{Q}) = \big\{U \subset \mathcal{Q} \big| \exists \text{ local timeshift } \mathcal{T}\colon\ U \in \mathcal{B}^*\big(\iota_{\mathcal{T}}(\mathcal{S}_0)\big)\big\}\,. \tag{3.20b}$$

◇

The next result formalizes the statement that the kinematic Born rule in the Eulerian picture is consistent with the instantaneous Born rule from Sect. 2. By Eq. (3.19i), it is, therefore, also consistent with the Born rule in the Lagrangian picture.

**Lemma 3.3** *Let $(\mathcal{Q}, g, \mathcal{O})$ be a spacetime, let $V$ be a future-directed timelike/causal/lightlike vector field on $\mathcal{Q}$ that is nowhere tangent to a hypersurface $(\mathcal{S}_0, \iota_0)$ in $\mathcal{Q}$. Furthermore, let $(\mathcal{S}, \iota)$ be the respective (maximal) 1-body flowout (cf. Theorem 3.2.(1).(i)) and let $\mathcal{T}$ be a global timeshift of $\mathcal{S}_0$ in $\mathcal{S}$. Define $\rho$ and the probability space $(\mathcal{S}_{\mathcal{T}}, \mathcal{B}^*(\mathcal{S}_{\mathcal{T}})\,, \mathbb{P}_E)$ as in Corollary 3.1.(4) above (so that Eq. (3.18) holds).*

*Then, the vector field $J = \rho V$ is future-directed timelike/causal/lightlike on $\mathcal{Q}$ and defines a vector field $J_{\mathcal{T}} = J_{\iota_{\mathcal{T}}(\,.\,)}$ over $(\mathcal{S}_0, \iota_{\mathcal{T}})$ that is nowhere tangent.*

*Moreover, if one defines the measure $\mathbb{P}_{\mathcal{T}}$ for $J_{\mathcal{T}}$ in analogy to Theorem 2.1.(3) above, then for all $U \in \mathcal{B}^*\big(\mathcal{S}_{\mathcal{T}}\big)$ one has*

$$\mathbb{P}_{\mathcal{T}}\,(U) = \mathbb{P}_E\,(U)\,. \tag{3.21}$$

*Thus, $\mathbb{P}_{\mathcal{T}}$ is a probability measure on $(\mathcal{S}_{\mathcal{T}}, \mathcal{B}^*(\mathcal{S}_{\mathcal{T}}))$.* ◇

Note that Eqs. (3.19i) and (3.21) also hold in the absence of probability conservation, if interpreted accordingly.

Due to Lemma 3.3, respective coordinate expressions for $\mathbb{P}_{\mathrm{E}}$ may be obtained from Lemma 2.1. Local coordinate expressions for the divergence of a vector field may be found in the literature (see e.g. Equation 3.4.10 in Ref. [208]).

We shall discuss an example in the Eulerian picture by continuing Examples 3.1 and 3.2 above.

*Example 3.3* Since $V$ is given by Eqs. (3.5a) and (3.5b), $\nu$ has values

$$\nu_{(ct,x,y)} = \sqrt{1 + \omega^2(x^2 + y^2)/c^2}\, \mathrm{d}x \wedge \mathrm{d}y + \omega x\, \mathrm{d}t \wedge \mathrm{d}x + \omega y\, \mathrm{d}t \wedge \mathrm{d}y. \tag{3.22a}$$

The values of $\rho$ are computed from Eq. (3.16c):

$$\rho(ct, x, y) = \frac{1}{2\pi\sigma^2\sqrt{1+\omega^2(x^2+y^2)/c^2}}\, e^{-(x^2+y^2)/2\sigma^2}. \tag{3.22b}$$

The function $\rho$ is independent of $t$, because the Gaussian is centered at the origin and rotationally symmetric. Hence, this property results out of our particular choice of $\varrho$ in Example 3.1 and is, therefore, not generic.

Finally, we shall consider the Born rule on the past light cone at 0 in $\mathbb{R}^3$. A calculation reveals that

$$\mathcal{T}(x_0, y_0) = -\frac{r_0}{\sqrt{c^2+\omega^2 r_0^2}} \tag{3.22c}$$

would define an appropriate global timeshift $\mathcal{T}$—if the function were smooth at 0. The closed set $\{0\} \subset \mathcal{S}_0$ has, however, measure zero, and it is therefore of no concern. Therefore, in accordance with Corollary 3.1.(5), for a set $U' \subseteq \mathcal{S}_0 \setminus \{0\}$ the probability that the body is located in the region $U = \iota_{\mathcal{T}}(U')$ is given by

$$\mathbb{P}_{\mathrm{E}}(U) = \int_{U'} \frac{1}{2\pi\sigma^2}\, e^{-(x^2+y^2)/2\sigma^2}\, \mathrm{d}x\, \mathrm{d}y \tag{3.22d}$$

◇

We conclude with a remark on rescaling the vector field $V$.

*Remark 3.6* As noted in the introduction of this section, the kinematic 1-body Born rule is invariant under rescalings of the vector field $V$:

If $f$ is a (smooth) strictly positive function on $\mathcal{Q}$, then $V' = fV$ is also a future-directed timelike/causal/lightlike vector field and $\rho' = \rho/f$ defines an invariant probability so that the respective current density vector field $J$, as given by Eq. (3.3), remains unchanged under the transformation. Therefore, $\mathbb{P}_{\mathrm{E}}$ and, accordingly, $\mathbb{P}_{\mathrm{L}}$ is invariant under the transformation—independent of whether one has probability conservation or not.

If one does have probability conservation, however, then the Eulerian continuity equation, Eq. (3.18), also holds true with respect to $\rho'$ and $V'$. There is indeed an underlying geometric reason for what may otherwise seem like a trivial calculation, namely one may view this fact as a consequence of the absolute invariance of $\rho\, \nu$ with respect to $V$ (cf. Corollary 3.1.(3)):

$$\mathcal{L}_{fV}(\rho\, \nu) = \mathcal{L}_{V'}(\rho'\, \nu') = 0. \tag{3.23}$$

The reader is referred to Proposition 4.2.16 in Ref. [177] for a version of the respective Poincaré–Cartan Theorem.

◇

## 4 Example: a single particle decays into vacuum

The purpose of this section is to illustrate by means of a simple example that the above theory is not necessarily limited to the case of a constant number of bodies. As such the goal is more of pedagogical nature than to construct a physically acceptable model. The reader is also referred to Sec. 6 in Ref. [169] for an analog in the non-relativistic theory. There it is shown that an analogous modification of the 1-body Schrödinger theory leads to a non-linearity in the respective modified Schrödinger equation (cf. Sect. 1.4 above). The central theorems of Ref. [171] may also be of interest.

The example we wish to consider is that of a single particle decaying into vacuum. As it is easiest to construct this example in the Lagrangian picture, we assume that a future-directed timelike/causal/lightlike 1-body flowout $(\mathcal{S}, \iota)$ is given for a hypersurface $(\mathcal{S}_0, \iota_0)$ in a spacetime $(\mathcal{Q}, g, \mathcal{O})$. Without loss of generality, we require that $\iota$ is surjective.

We shall assume that initially there is always one particle in the ensemble, i.e. the Lagrangian probability density $\varrho$ with $\varrho_0 = \varrho(0, \,.\,)$ satisfies $\mathbb{P}_\mathrm{L}(\mathrm{graph}(0)) = 1$, and that for subsequent times $\varrho$ satisfies the differential equation

$$\mathring{\varrho} = -k\varrho \tag{4.1}$$

for some 'rate constant' $k > 0$. Taking local coordinates as in Lemma 3.2, we obtain the solution

$$\varrho\,(\tau, \xi)\,\sqrt{-\det(\iota^* g)}(\tau, \xi) = \varrho_0\,(\xi)\,\sqrt{-\det(\iota^* g)}(0, \xi)\;e^{-k\tau}. \tag{4.2}$$

Though this may seem like an oversimplified model and we have to restrict its validity to $\tau \geq 0$, the function $\rho = \varrho \circ \iota^{-1}$ might be quite complicated regardless. Due to Eqs. (3.19c) to (3.19e), the latter satisfies the partial differential equation

$$\mathrm{div}\,(\rho\,V) = -k\rho. \tag{4.3}$$

Of course, in an actual physical model, the solution in Eq. (4.2) would only be of limited use, since one would generally need to solve for $\varrho$ and $\iota$ (or $\rho$ and $V$, or even just $J$) simultaneously.

Still, Eq. (4.2) shows that the density $\varrho(\tau, \,.\,)$ tends to the zero function as $\tau \to \infty$, so that the probability to find any particle at all likewise goes to zero in this limit.

While the position probability is, therefore, not conserved, it is still possible to speak of probability conservation in a generalized sense by introducing the quantity

$$\Theta_\mathcal{T} = 1 - \mathbb{P}_\mathrm{E}\,(\mathcal{S}_\mathcal{T}) \tag{4.4}$$

for any (here positive) global timeshift $\mathcal{T}$ of $(\mathcal{S}_0, \iota_0)$ in $(\mathcal{S}, \iota)$. The quantity $\Theta_\mathcal{T}$ gives the probability that no particles are located in the hypersurface $\mathcal{S}_\mathcal{T}$—i.e. no particles 'anywhere' (vacuum) at this 'instance of time'.

Probability conservation in this generalized sense now means that the probability that there is either no or one particle is 1 for any choice of time shift.

We once again conclude with a remark.

*Remark 4.1* (1) One of the challenges of formulating continuity equations with sources, like the ones in Eqs. (4.1) and (4.3), is that the quantity $\mathbb{P}_\mathrm{L}(\mathrm{graph}(\mathcal{T})) = \mathbb{P}_\mathrm{E}\,(\mathcal{S}_\mathcal{T})$ has to remain in the interval [0, 1]. In formulating such source terms one may take inspiration from chemical kinetics and the theory of reaction-(advection-)diffusion systems, where the amount of substance also has to be bounded from below and above. See e.g. Problem 10 in Sec. 15.2 of Ref. [134] for an example of the latter (without advection).[65]

(2) One may in principle formulate 'higher-order continuity equations' in the Lagrangian picture by using $\varrho$, $\mathring{\varrho}$, $\mathring{\mathring{\varrho}}$, etc. as well as other functions on $\mathcal{S}$. Using coordinates as in Lemma 3.2, we find, for instance, that

$$\mathring{\mathring{\varrho}}(\tau, \xi) = \frac{1}{\sqrt{-\det(\iota^* g)}(\tau, \xi)}\,\frac{\partial^2}{\partial \tau^2}\left(\varrho(\tau, \xi)\,\sqrt{-\det(\iota^* g)}(\tau, \xi)\right). \tag{4.5}$$

Of course, the use of higher order derivatives also requires either more initial conditions or more partial differential equations for the entire initial value problem to be solvable.

[65] Unfortunately, the treatments of diffusion systems in the literature that we are aware of do not explicitly spell out the underlying continuity equations. The latter can, however, be reobtained from the respective diffusion equation by an analysis of the currents via Fick's first law.

(3) When considering continuity equations with source terms in this theory it is important to keep in mind that the function $\varrho$ in the Langrangian picture has to be nowhere-vanishing.
While the condition may seem superfluous, given that the Born rule makes sense regardless, there is the problem that $\varrho$ and $\iota$ are in general dynamically coupled by respective partial differential equations. Roughly speaking, the map $\iota$ cannot propagate probability that is not there to begin with. A similar issue occurs in the non-relativistic theory, where the drift field $\vec{v}$ is ill-defined in regions where the (suitably regular) wave function vanishes.
In the relativistic theory and for the case that one has probability conservation, the problem may be dealt with by restricting oneself to a domain of $\varrho_0 = \varrho(0, \, . \,)$ on which it does not vanish. If one does not have probability conservation, then despite this restriction it is straightforward to construct an example for which $\varrho$ vanishes entirely on some graph of a global timeshift. For the aforementioned physical reason, such behavior is to be excluded.

◇

**Acknowledgements** The authors would like to acknowledge support from The Robert A. Welch Foundation (D-1523). M.R. also acknowledges financial support from the Department of Mathematics and Statistics at Texas Tech University. Furthermore, M. R. thanks Salvador Miret-Artés, Yuri B. Suris, and Gerd Rudolph for their support in making this research possible. M. R. also thanks Markus Fenske for his IT assistance, Hung Tran for his help with sharing the results of this work with a wider audience, Joseph D. Romano for writing advice, and Mike Scherfner for constructive discussions. Roderich Tumulka deserves gratitude for pointing out Ref. [151] and for helpful criticism of a prior version of this work. Finally, both authors would like to thank the anonymous referee and the managing editor, Lev Vaidman, for their constructive comments, which helped to improve the manuscript.

**Funding** Open Access funding enabled and organized by Projekt DEAL.

**Data availability** The numerical data generated for the purpose of this study is exhibited in Fig. 4.

**Declarations**

**Conflict of interest** The authors do not have any conflict of interest to declare.

## Appendix

### Appendix A Proofs

*Proof of Theorem 2.1*

(1) This is a corollary of Prop. 15.21 in Ref. [127].
(2) By footnote 47 the expression in Eq. (2.2) is a well-defined and smooth top-degree form on $\mathcal{S}_0$.
In order to show that it is nowhere vanishing, let $(Y_1, \ldots, Y_n)$ be any oriented frame at $q$ in $\mathcal{S}_0$. We find that

$$\left(\iota_0^*\left(J_0 \lrcorner \, \mu_{\iota_0(\,.\,)}\right)\right)_q (Y_1, \ldots, Y_n) = \mu_{\iota_0(q)} \left((J_0)_q, (\iota_0)_* Y_1, \ldots, (\iota_0)_* Y_n\right). \tag{A.1}$$

The right hand side of Eq. (A.1) is non-vanishing, since the non-zero vector $(J_0)_q$ is transverse to the subspace $(\iota_0)_* \, \mathrm{T}_q \mathcal{S}_0$ of $\mathrm{T}_q \mathcal{Q}$ and $\mu$ is nowhere vanishing. Thus, the expression in Eq. (2.2) is nowhere vanishing. Since the orientation in point (1) was chosen such that the right hand side of Eq. (A.1) is positive, we indeed have a volume form on $\mathcal{S}_0$.

(3) Due to the normalization, Eq. (2.4a), and point (2), $\mathbb{P}_0$ does indeed take values in [0, 1], provided $\iota_0^{-1}(U)$ is measurable. Since $(\mathcal{S}_0, \iota_0)$ and $\iota_0(\mathcal{S}_0)$ (together with the natural inclusion) are equivalent submanifolds, $\iota_0^{-1}(U)$ is indeed measurable whenever $U$ is measurable.
For the second statement we use this equivalence to simplify our notation, thus not distinguishing between $\mathcal{S}_0$ and $\iota_0(\mathcal{S}_0)$ and their respective subsets:
To show the reverse implication, recall that $U$ is a Lebesgue null set if and only if it intersects to a null set in each chart $(V, \xi)$ (cf. Prop. 1.6 in Chap. XII of Ref. [3]). Furthermore, observe that point (2) implies that there exists a strictly positive scalar field $\chi$ on $\xi(V) \subseteq \mathbb{R}^n$ such that the following coordinate expression holds

$$\iota_0^* \left(J_0 \lrcorner \mu_{\iota_0(\,.\,)}\right) = \chi \, \mathrm{d}^n \xi. \tag{A.2}$$

Therefore, the measure $U \mapsto \mathbb{P}_{\mathcal{S}_0}(V \cap U)$ is absolutely continuous with respect to the Lebesgue measure on $V$ induced by the coordinates $\xi$ (cf. Def. 7.30 in Ref. [117]). To complete the proof, choose an at most countable atlas $(V_\alpha, \xi_\alpha)_{\alpha \in I}$ for $\mathcal{S}_0$. Then we find that

$$\mathbb{P}_{\mathcal{S}_0}(U) = \mathbb{P}_{\mathcal{S}_0}\left(\left(\bigcup_{\alpha \in I} V_\alpha\right) \cap U\right) \leq \sum_{\alpha \in I} \mathbb{P}_{\mathcal{S}_0}(V_\alpha \cap U) = 0. \tag{A.3}$$

For the forward implication, take $(V, \xi)$ as above and use the fact that

$$0 = \mathbb{P}_{\mathcal{S}_0}(U) \geq \mathbb{P}_{\mathcal{S}_0}(U \cap V) \geq 0. \tag{A.4}$$

From Eq. (A.2) and strict positivity of $\chi$ we conclude that $U \cap V$ has Lebesgue measure zero (cf. Ex. 7.32 (i) in Ref. [117]). Since this holds for any chart $(V, \xi)$, $U$ itself is a null set.

*Proof of Proposition 2.1*

(2) Pick any $q \in \mathcal{Q}$. By assumption, $J_q \neq 0$. By the straightening lemma, there exist local coordinates $\kappa$ in an open neighborhood $U$ of $q$ such that $J = \partial/\partial\kappa^0$ on $U$ (cf. Prop. 3.2.17.2 in Ref. [177]). Now choose $\mathcal{S}_0$ to be the $\kappa^0 = 0$ slice.
(1) Due to point (2), it is enough to show that locally there always exists a future-directed timelike/causal/lightlike vector field $J$ on $\mathcal{Q}$. For the timelike case we refer to Lem. 5.32 and Prop. 5.37 in Ref. [153]. For the lightlike case, Lemma B.1 from Appendix B is sufficient. The causal case then follows as well.

*Proof of Lemma 2.1*

(2) The proof uses the fact that one can 'shift around' the anti-symmetrization operation, if each tensor component is a product of tensor components of type (1, 1):

$$\kappa^* \left(J_0 \lrcorner \mu_{\kappa(\,.\,)}\right) = \left(\sqrt{-\det g} \circ \kappa\right) \frac{1}{n!} \varepsilon_{i_0 \dots i_n} J_0^{i_0} \, \mathrm{d}\kappa^{i_1} \wedge \dots \wedge \mathrm{d}\kappa^{i_n} \tag{A.5}$$

$$= \left(\sqrt{-\det g} \circ \kappa\right) \frac{1}{n!} \varepsilon_{i_0 \dots i_n} J_0^{i_0} \frac{\partial \kappa^{i_1}}{\partial \xi^{j_1}} \dots \frac{\partial \kappa^{i_n}}{\partial \xi^{j_n}} \mathrm{d}\xi^{j_1} \wedge \dots \wedge \mathrm{d}\xi^{j_n} \tag{A.6}$$

$$= \left(\sqrt{-\det g} \circ \kappa\right) \varepsilon_{i_0 \dots i_n} J_0^{i_0} \frac{\partial \kappa^{i_1}}{\partial \xi^{[1}} \dots \frac{\partial \kappa^{i_n}}{\partial \xi^{n]}} \mathrm{d}\xi^{1} \wedge \dots \wedge \mathrm{d}\xi^{n} \tag{A.7}$$

$$= \left(\sqrt{-\det g} \circ \kappa\right) \varepsilon_{i_0 \dots i_n} J_0^{i_0} \frac{\partial \kappa^{[i_1}}{\partial \xi^{1}} \dots \frac{\partial \kappa^{i_n]}}{\partial \xi^{n}} \mathrm{d}\xi^{1} \wedge \dots \wedge \mathrm{d}\xi^{n} \tag{A.8}$$

$$= \left(\sqrt{-\det g} \circ \kappa\right) \varepsilon_{i_0 [i_1 \dots i_n]} J_0^{i_0} \frac{\partial \kappa^{i_1}}{\partial \xi^{1}} \dots \frac{\partial \kappa^{i_n}}{\partial \xi^{n}} \mathrm{d}\xi^{1} \wedge \dots \wedge \mathrm{d}\xi^{n} \tag{A.9}$$

$$= \left(\sqrt{-\det g} \circ \kappa\right) \, \varepsilon_{i_0 i_1 \ldots i_n} \, J_0^{i_0} \frac{\partial \kappa^{i_1}}{\partial \xi^1} \ldots \frac{\partial \kappa^{i_n}}{\partial \xi^n} \, \mathrm{d}\xi^1 \wedge \ldots \wedge \mathrm{d}\xi^n \tag{A.10}$$

$$= \left(\sqrt{-\det g} \circ \kappa\right) \, (n+1)! \, J_0^{[0} \frac{\partial \kappa^1}{\partial \xi^1} \ldots \frac{\partial \kappa^{n]}}{\partial \xi^n} \, \mathrm{d}\xi^1 \wedge \ldots \wedge \mathrm{d}\xi^n . \tag{A.11}$$

(1) We take the expression from Eq. (A.10) and observe that

$$\varepsilon_{i_0 i_1 \ldots i_n} \, J_0^{i_0} \, \delta_1^{i_1} \, \ldots \, \delta_n^{i_n} = \varepsilon_{0\,1\ldots n} \, J_0^0 = J_0^0 . \tag{A.12}$$

*Proof of Proposition 3.1*

The map $\Phi_\tau$ is a diffeomorphism from dom $\Phi_\tau$ to dom $\Phi_{-\tau}$ (cf. Thm. 9.12(c) in Ref. [127] or Prop. 3.2.10.1 in Ref. [177]). Therefore the map $\iota(\tau, \,.\,) = \Phi_\tau \circ \iota_0$ is a smooth embedding with respect to the smooth structure on dom $\Phi_{-\tau}$. Yet dom $\Phi_{-\tau}$ is an open submanifold of $\mathcal{Q}$, so we may view $\iota(\tau, \,.\,)$ as a smooth embedding to $\mathcal{Q}$. Therefore $\mathcal{S}_\tau$ indeed admits a smooth structure such that, together with its natural inclusion, it becomes an embedded submanifold. The respective smooth structure is unique (cf. Thm. 3.51 in Ref. [127]).

To check the transversality condition, observe that for any $q \in \iota_0(\mathcal{S}_0)$ a vector $Y$ in $\mathrm{T}_{\Phi_\tau(q)}\mathcal{Q}$ is tangent to $\mathcal{S}_\tau$ if and only if $(\Phi_{-\tau})_* Y$ is tangent to $(\mathcal{S}_0, \iota_0)$. Yet any vector field $V$ is invariant under its own flow, so that

$$(\Phi_{-\tau})_* \left(V_{\Phi_\tau(q)}\right) = V_q \, . \tag{A.13}$$

The assertion now follows from the fact that $V_q$ is transverse to $(\mathcal{S}_0, \iota_0)$.

*Proof of Proposition 3.2*

(1) We first show the timelike case:
Since $\mathcal{Q}$ is time-oriented (via the spacetime-orientation $\mathcal{O}$), we may again choose a global future-directed timelike vector field $V$ (cf. Lem. 5.32 and Prop. 5.37 in Ref. [153]). Further, by Proposition 2.1, we may also choose an orientable hypersurface $(\mathcal{S}_0, \iota_0)$ that is nowhere tangent to $V$.
By applying the flowout theorem (Thm. 9.20 in Ref. [127]), we obtain a tuple $(\mathcal{S}, \iota)$ that satisfies all properties of a future-directed timelike 1-body flowout provided that $\iota$ is a topological embedding. As stated in point (d) of Thm. 9.20 in Ref. [127], this missing property follows from dimensional considerations (cf. Prop. 4.22(d) in Ref. [127]).
For the lightlike case we employ Lemma B.1 in Appendix B to obtain a future-directed lightlike vector field $V$ on some open subset $U$ of a spacetime $\mathcal{Q}$. Then, as an open submanifold of $\mathcal{Q}$, the set $U$ is canonically a spacetime as well, so we may apply Proposition 2.1 to obtain a suitable hypersurface $\mathcal{S}_0$ in $U$. The assertion then follows by repeating the argument for the timelike case.
The causal case follows from the timelike or the lightlike case.
(2) This is a consequence of the flowout theorem, as argued in point (1).
(3) Existence of a smooth extension $V$ of $V_0$ on some open neighborhood $U$ of $\iota_0(\mathcal{S}_0)$ follows from the fact that $(\mathcal{S}_0, \iota_0)$ is embedded.[66] Moreover, since the set $\left(g(V, V)\right)^{-1}(\mathbb{R}_+)$ is open in $\mathcal{Q}$ and contains $\mathcal{S}_0$, we may choose $U$ such that $V$ is timelike on $U$. Since $V_0$ is future-directed over $\mathcal{S}_0$ and $V$ is continuous, $V$ is future-directed on $U$. The assertion now follows from point (2).

[66] See Problem 8–15 in Ref. [127]. The proof is essentially a partition of unity argument.

*Proof of Lemma 3.1*

The image $\iota(\mathcal{S})$ is an open submanifold of $\mathcal{Q}$ and therefore defines a spacetime by restricting $g$ and $\mathcal{O}$ as stated. Now, since $\tilde{\iota}$ is a diffeomorphism, it gives rise to spacetime $\mathcal{S}$ isomorphic to $\iota(\mathcal{S})$.

*Proof of Theorem 3.1*

(1) (i) See Prop. 5.4 in Ref. [127] and its proof.
 (ii) Since $\dim \mathcal{S} = \dim \mathcal{Q}$ and $\iota$ is an immersion, $\iota^*\mu$, as given in point (2), is a nowhere vanishing top-degree form on $\mathcal{S}$. We define a frame $(Y_1, \ldots, Y_n)$ at $q$ in $\mathcal{S}_0$ to be oriented, if

$$\left((\mathcal{T}')^* \upsilon\right)_q (Y_1, \ldots, Y_n) = \frac{1}{c}\, \mu_{\iota\mathcal{T}(q)} \left(\iota_* \left(\left.\frac{\partial}{\partial \tau}\right|_{\mathcal{T}'(q)}\right), (\iota_\mathcal{T})_* Y_1, \ldots, (\iota_\mathcal{T})_* Y_n\right) \tag{A.14}$$

is greater than zero. The orientation is well-defined, because $\partial/\partial\tau$ is nowhere tangent to graph$(\mathcal{T})$. It is canonical, because $\mathcal{T}'$ is an orientation-preserving map whenever $\mathcal{S}_0$ is equipped with the orientation induced by the vector field $\mathcal{V}_{\mathcal{T}'(\,.\,)}$ over $(\mathcal{S}_0, \iota_\mathcal{T})$ (cf. Prop. 15.21 in Ref. [127]).

(2) This follows from the definition of the orientation in the proof of point (1).(ii).

(3) Since $\varrho$ is strictly positive and $\upsilon$ is nowhere vanishing, $\varrho\, \upsilon$ is nowhere vanishing. Identity (3.12c) is now a trivial consequence of the definition of the derivative on the left hand side (cf. Footnote 62). Uniqueness follows from the fact that, if $(h - h')\, \upsilon = 0$ for $h, h' \in C^\infty(\mathcal{S}, \mathbb{R})$, then $h = h'$.

(5) Due to $\mathring{\varrho} = 0$ and Eq. (3.12c), $\varrho\, \upsilon$ does not depend on $\tau$. That is, for all $(\tau, q) \in \mathcal{S}$ we have[67]

$$\varrho(\tau, q)\, \upsilon_{(\tau,q)} = \varrho(0, q)\, \upsilon_{(0,q)}. \tag{A.15}$$

Since $\mathcal{T}' = (\mathcal{T}, \mathbb{1}_{\mathcal{S}_0})$, we conclude that

$$(\mathcal{T}')^* (\varrho\, \upsilon) = \varrho(0,\,.\,)\, \upsilon_{(0,\,.\,)}. \tag{A.16}$$

Finally, observe that

$$\int_U \varrho\, \upsilon = \int_{(\mathcal{T}')^{-1}(U)} (\mathcal{T}')^* (\varrho\, \upsilon) = \int_{\mathrm{pr}_2(U)} \varrho(0,\,.\,)\, \upsilon_{(0,\,.\,)}. \tag{A.17}$$

(4) The first point follows from Eq. (A.17) and the normalization (3.12d). For the second point, argue as in the proof of Theorem 2.1.(3).

*Proof of Lemma 3.2*

(2) Using the general coordinate expression for volume forms induced by a metric, we have

$$\left(\iota^* \mu\right) = \sqrt{-\det\left(\iota^* g\right)}\, \mathrm{d}\tau \wedge \mathrm{d}\xi^1 \wedge \ldots \wedge \mathrm{d}\xi^n. \tag{A.18}$$

The integrand is obtained by contracting with $\partial/\partial\tau$ and multiplying by $\varrho/c$.

(1) One way to show this is to realize that $\mathring{\varrho}$ is a divergence, which allows one to use the respective formula. Another approach is to derive the integrand from point 3.15 directly and compare the two sides of Eq. (3.12c).

[67] Though $\upsilon_{(\tau,q)}$ and $\upsilon_{(0,q)}$ are formally elements of different vector spaces, the equation is sensible, if we view $\tau$ as a parameter (cf. Footnote 62). The latter is possible because $\upsilon$ does not have any ‘d$\tau$ components’.

*Proof of Theorem 3.2*

(1) (i) See Proposition 3.2.(2). Equations (3.19a) and (3.19b) hold by construction.
(ii) This is also true by construction.

(2) Set $n = \dim \mathcal{S}_0$. Then for all $(\tau, q) \in \mathcal{S}$ and $Y_1, \ldots, Y_n \in \mathrm{T}_{(\tau,q)}\mathcal{S}$ we have

$$\left(\iota^* \nu\right)_{(\tau,q)} (Y_1, \ldots, Y_n) = \frac{1}{c}\, \mu_{\iota(\tau,q)} \left(V, \iota_* Y_1, \ldots, \iota_* Y_n\right) \tag{A.19}$$

$$= \frac{1}{c}\, \mu_{\iota(\tau,q)} \left(\iota_* \frac{\partial}{\partial \tau}, \iota_* Y_1, \ldots, \iota_* Y_n\right) \tag{A.20}$$

$$= \upsilon_{(\tau,q)} \left(Y_1, \ldots, Y_n\right). \tag{A.21}$$

(3) The respective functions are well defined in either case.
We first show Eq. (3.17c) using Cartan's formula (cf. Prop. 4.1.8 in Ref. [177]): On one hand, by the definition of the divergence (cf. Footnote 64), we have

$$\mathcal{L}_{(\rho\, V)} \mu = \mathrm{d} \left(\rho\, V \lrcorner\, \mu\right) = \mathrm{div} \left(\rho\, V\right) \mu. \tag{A.22}$$

But then Eq. (3.17c) follows from

$$\mathcal{L}_V \left(\rho\, V \lrcorner\, \mu\right) = V \lrcorner\, \mathrm{d} \left(\rho\, V \lrcorner\, \mu\right) \tag{A.23}$$

$$= \mathrm{div} \left(\rho\, V\right)\; V \lrcorner\, \mu. \tag{A.24}$$

Using an identity for the behavior of the Lie derivative under diffeomorphisms (cf. Prop. 3.3.3.5 and Eq. 2.5.7 in Ref. [177]), Eq. (3.17c) as well as the definition in Footnote 62, we may now show Eq. (3.19e):

$$\mathring{\varrho}\, \upsilon = \mathcal{L}_{\partial/\partial\tau} \big( (\rho \circ \iota)\, \iota^* \nu \big) \tag{A.25}$$

$$= \mathcal{L}_{(\iota^{-1})_* V} \big( \iota^* (\rho\, \nu) \big) \tag{A.26}$$

$$= \iota^* \big( \mathcal{L}_V (\rho\, \nu) \big) \tag{A.27}$$

$$= \iota^* \big( \mathrm{div}(\rho\, V)\, \nu \big). \tag{A.28}$$

Equation (3.19f) is a trivial—yet important—consequence thereof.

(4) '(i) $\Longrightarrow$ (ii)': Since $\iota$ is a diffeomorphism, property (i) holds if and only if for every such $q$ the (maximal) integral curve $\tau \mapsto (\tau, q)$ of $\partial/\partial\tau$ intersects the hypersurface $\iota^{-1}\big(\varphi(\mathcal{N})\big)$ in $\mathcal{S} \subseteq \mathbb{R} \times \mathcal{S}_0$ transversally in exactly one point. Existence and smoothness of $\mathcal{T}$ such that

$$\iota^{-1}\big(\varphi(\mathcal{N})\big) = \mathrm{graph}(\mathcal{T}) \tag{A.29}$$

now follows from Thm. 6.32 in Ref. [127].
'(ii) $\Longrightarrow$ (i)': This holds by definition of $\iota$ and the transversality condition.

(5)

$$\int_{\iota(\mathcal{T}'(U))} \rho\, \nu = \int_{\mathcal{T}'(U)} \iota^* \left(\rho\, \nu\right) = \int_{\mathcal{T}'(U)} \varrho\, \upsilon. \tag{A.30}$$

*Proof of Corollary 3.1*

Once one identifies the hypersurface $(\mathcal{N}, \varphi)$ with $(\mathcal{S}_0, \iota_{\mathcal{T}})$ using Theorem 3.2.(4), Corollary 3.1 is mostly a consequence of Theorem 3.1. Upon having made this identification, the more detailed arguments are as follows:

(1) Point (i) is trivial. Point (ii) follows from Theorem 3.1.(1).(ii) and the fact that $\iota$ is an injective immersion.
(2) See Theorem 3.1.(2) and Theorem 3.2.(2).
(3) Consult the proof of Theorem 3.2.(3) above. Absolute invariance follows from its definition together with the fact that $V \lrcorner (V \lrcorner \mu)$ vanishes due to linear dependence of the two contractions.
(4) Theorem 3.2.(3) states that $\mathring{\varrho} = 0$ in this instance. The set $V = \iota_{\mathcal{T}}^{-1}(U)$ is a Lebesgue set and so is $\mathcal{T}'(V) \subseteq \mathrm{graph}(\mathcal{T})$. The assertion now follows from combining Theorem 3.2.(5) and 3.1.(4).
(5) We apply Theorem 3.2.(5), Theorem 3.1.(5), and then the former again:

$$\mathbb{P}_{\mathrm{E}}\big(\iota_{\mathcal{T}}(U)\big) = \mathbb{P}_{\mathrm{L}}\big(\mathcal{T}'(U)\big) = \mathbb{P}_{\mathrm{L}}\big(0'(U)\big) = \mathbb{P}_{\mathrm{E}}\big(\iota_0(U)\big). \tag{A.31}$$

*Proof of Lemma 3.3*

$J$ is future-directed timelike/causal/lightlike, since this is the case for $V$ and $\rho$ is strictly positive. $J_{\mathcal{T}}$ is shown to be nowhere tangent to $(\mathcal{S}_0, \iota_{\mathcal{T}})$ by adapting the second part of the proof of Proposition 3.1 (set $\tau = \mathcal{T}(q)$). Equation (3.21) holds by construction.

## Appendix B Local existence of future-directed lightlike vector fields

The following lemma is used in the proof of Proposition 3.2.

**Lemma B.1** *Locally, every time-oriented Lorentzian manifold admits a future-directed lightlike (smooth) vector field.* ◇

*Proof* Since $\mathcal{Q}$ is time-oriented, we may choose a global future-directed timelike vector field $V$ (cf. Lem. 5.32 and Prop. 5.37 in Ref. [153]). Then, by local triviality of the respective principal bundle, for every $q \in \mathcal{Q}$ there exists an open neighborhood $U$ of $q$ in $\mathcal{Q}$ and a time-oriented orthonormal frame field $Z$ over $U$. Without loss of generality, we may assume that $Z_0$ is parallel to $V$ on $U$. Further, if $\Lambda \in \mathrm{GL}\left(\mathbb{R}^4\right)$ is any non-trivial Lorentz boost, then the vector field

$$Y = (Z \cdot \Lambda)_0 \tag{B.1}$$

is future-directed timelike, normalized, and nowhere parallel to $V$. One shows that the parallel and orthogonal components of $V$ with respect to $Y$, denoted by $V^{\parallel}$ and $V^{\perp}$, are smooth and nowhere vanishing. Upon defining $K = V^{\parallel} + \lambda V^{\perp}$ with

$$\lambda = \sqrt{g\left(V^{\parallel}, V^{\parallel}\right) / - g\left(V^{\perp}, V^{\perp}\right)}, \tag{B.2}$$

we find that $K$ is lightlike. Moreover, $K$ is smooth.

To show that $K$ is future-directed, consider the sequence $(K_n')_{n\in\mathbb{N}}$ with

$$K_n' = V^{\parallel} + \left(1 - \frac{1}{n}\right) \lambda V^{\perp}. \tag{B.3}$$

By linear independence of the nowhere vanishing vector fields $V^{\parallel}$ and $V^{\perp}$, every $K_n'$ is nowhere vanishing. Furthermore, $K_1' = V^{\parallel}$ is future-directed timelike and $g(K_n', K_n')$ is strictly positive for all $n \in \mathbb{N}$. Therefore, for every $q \in U$ the sequence $\left(\left(K'_n\right)_q\right)_{n\in\mathbb{N}}$ stays in the (closed) future tangent light cone at $q$ and converges to a lightlike vector $K_q$. Thus $K$ is indeed future-directed. □

## Appendix C Generalized flowouts

As noted in Remark 3.1.(3), there is reason to generalize Definition 3.2 so that it adequately accounts for periodic integral curves. We shall provide such an intrinsic definition here and also prove a theorem establishing its suitability.

**Definition C.1** Let $\mathcal{Q}$ be a manifold and let $(\mathcal{S}_0, \iota_0)$ be an embedded submanifold.

*A flowout from* $(\mathcal{S}_0, \iota_0)$ is a tuple $(\mathcal{S}, \iota)$ such that

(1) $\mathcal{S}$ is a flow domain for $\mathcal{S}_0$, and

(2) the map

$$\iota\colon \mathcal{S} \to \mathcal{Q}\colon (\tau, q) \mapsto \iota(\tau, q) \tag{C.1a}$$

satisfies all of the properties below:

(i) $\iota(0,\ .\,) = \iota_0$ .

(ii) $\iota$ is an immersion.

(iii) For all $(\tau, q)$ and $(\tau', q')$ in $\mathcal{S}$ the following implication holds:

$$\iota(\tau, q) = \iota(\tau', q') \quad \Longrightarrow \quad q = q' \ \text{ and } \ \iota_* \left.\frac{\partial}{\partial \tau}\right|_{(\tau,q)} = \iota_* \left.\frac{\partial}{\partial \tau}\right|_{(\tau',q')} . \tag{C.1b}$$

(iv) If the image $\iota(\mathcal{S})$ is equipped with the subspace topology in $\mathcal{Q}$, then the restriction of $\iota$ to $\iota(\mathcal{S})$ in codomain is a quotient map. ◇

Upon replacing Definition 3.2.(1) with Definition C.1, the definition of flowouts along vector fields and maximal flowouts is the same as in Definition 3.2.(2). Similarly, the formal definition of 1-body flowouts stays the same (cf. Definition 3.3).

Definition C.1 disposes of the assumptions that $\iota$ has to be injective and a topological embedding. One may understand points (2).(iii) and (2).(iv) as a weakening of the former condition and the latter topological condition, respectively. Definition C.1 does indeed provide a mathematical generalization of Definition 3.2. Hence, there is no need to prove a separate flowout theorem for this more general definition.[68]

In the particular case that $(\mathcal{S}_0, \iota_0)$ is a hypersurface, any immersion $\iota\colon \mathcal{S} \to \mathcal{Q}$ is a local diffeomorphism (cf. Prop. 4.8 in Ref. [127]). Therefore, it automatically satisfies the respective topological condition.

We proceed to show that the use of the term 'flowout' in Definition C.1 is indeed justified. The respective statement is given by Theorem C.1 below. The latter relies on the following lemma.

**Lemma C.2** *Let $\mathcal{S}$ and $\mathcal{Q}$ be smooth manifolds, and let $\iota$, as in Eq.* (C.1a)*, be an immersion that satisfies condition (2).(iv) of Definition* C.1.

*Then there exists a unique smooth structure on the image $\iota(\mathcal{S})$ such that (together with its natural inclusion) it is an embedded submanifold of $\mathcal{Q}$.* ◇

*Proof* The general idea of proof is to use the rank theorem to construct slice charts on $\iota(\mathcal{S})$ (cf. Thms. 4.12 and 5.8 in Ref. [127]), which is possible because of the topological properties of $\iota$.

Again, denote the restriction of $\iota$ in codomain by $\tilde{\iota}$ and the (necessarily injective) natural inclusion of $\iota(\mathcal{S})$ in $\mathcal{Q}$ by $\varphi$. Clearly, $\iota = \varphi \circ \tilde{\iota}$.

[68] Do note the point raised in Remark 3.1.(1), however.

Since $\iota$ is an immersion, we may indeed apply the rank theorem: For every $p \in \mathcal{S}$ there exists a smooth chart $(V, \zeta)$ on $\mathcal{S}$ around $p$ and a smooth chart $(U, \kappa)$ on $\mathcal{Q}$ around $\iota(p)$ such that $\iota(V) \subseteq U$ and $\iota$ has the local coordinate representation

$$\zeta \mapsto \kappa(\zeta) = \left(\zeta^1, \ldots, \zeta^n, 0, \ldots, 0\right) \tag{C.2}$$

with $n = \dim \mathcal{S}$.

By assumption, $\tilde{\iota}$ is an open map. Thus $\iota(V) = \tilde{\iota}(V)$ is open with respect to the subspace topology on $\iota(\mathcal{S})$ in $\mathcal{Q}$. Moreover, the restriction $\alpha_V$ of $\tilde{\iota}$ to $V$ in domain and to $\iota(V)$ in codomain is a homeomorphism. Therefore, $\left(\iota(V), \zeta \circ (\alpha_V)^{-1}\right)$ defines a chart on $\iota(\mathcal{S})$. The respective transition functions are smooth by construction, and thus the collection of all such charts gives rise to a smooth structure on $\iota(\mathcal{S})$.

Therefore, locally we may view the map $\tilde{\iota}$ as the identity mapping $\xi \mapsto \xi$. We may hence understand the map in Eq. (C.2) as a (smooth) local representative of $\varphi$.

Finally, we recall that $\iota(\mathcal{S})$ carries the subspace topology. Thus there exists an open $U'$ in $\mathcal{Q}$ such that $\iota(V) = U' \cap \iota(\mathcal{S})$. It follows that the coordinates $\kappa$ restrict to slice coordinates for $\iota(\mathcal{S})$ on $U' \cap U$. Thus, $(\iota(\mathcal{S}), \varphi)$ is an embedded submanifold of $\mathcal{Q}$, and as such, the choice of smooth structure is unique (cf. Thm. 5.31 in Ref. [127]). □

**Theorem C.1** *Let $(\mathcal{S}, \iota)$ be a flowout from an embedded submanifold $(\mathcal{S}_0, \iota_0)$ of a manifold $\mathcal{Q}$ (according to Definition C.1 above).*

*Then the following holds:*

(1) *There exists a unique smooth structure on $\iota(\mathcal{S})$ such that $\iota(\mathcal{S})$ (together with its natural inclusion) is an embedded submanifold of $\mathcal{Q}$.*
(2) *There exists a unique smooth vector field $V$ on $\iota(\mathcal{S})$ such that*

$$V_{\iota(\tau,q)} = \iota_* \left.\frac{\partial}{\partial \tau}\right|_{(\tau,q)} \tag{C.3}$$

*for all $(\tau, q) \in \mathcal{S}$.*
(3) *There exists a (non-unique) smooth extension $\tilde{V}$ of $V$ to an open neighborhood of $\iota(\mathcal{S})$ in $\mathcal{Q}$. In addition, $(\mathcal{S}, \iota)$ is a flowout from $(\mathcal{S}_0, \iota_0)$ along $\tilde{V}$.* ◇

*Proof* (1) This follows from Lemma C.2 above.
(2) $V$ is well-defined: If $p \in \iota(\mathcal{S})$, then there exists a $(\tau, q) \in \mathcal{S}$ such that $p = \iota(\tau, q)$. Now define $V_p$ via Eq. (C.3). Due to the second part of Eq. (C.1b), $V_p$ is indeed uniquely defined in this manner.
$V$ is smooth: Recall that, as shown in the proof of Lemma C.2, the natural inclusion $\varphi$ is locally given by Eq. (C.2). With respect to these coordinates, we have

$$V = \frac{\partial \kappa^i}{\partial \tau} \frac{\partial}{\partial \kappa^i}. \tag{C.4}$$

Smoothness of $V$ therefore follows from smoothness of $\varphi$ (or smoothness of $\iota$).
(3) We may extend $V$ using the extension lemma for vector fields on submanifolds (cf. Problem 8–15 in Ref. [127]). The second assertion holds by construction.

□

It is noteworthy that above we did not make any explicit use of the first point on the right hand side of the implication (C.1b). The condition is nonetheless justified by the fact that it will hold, whenever $\iota$ arises from the flow of a vector field (in an open neighborhood of $\iota(\mathcal{S})$) in $\mathcal{Q}$.

**Publisher's Note** Springer Nature remains neutral with regard to jurisdictional claims in published maps and institutional affiliations.